\documentclass[11pt]{article}
\RequirePackage{charter,fullpage,enumitem,authblk,natbib,amsthm,amsmath,amsfonts,amssymb,graphicx,subcaption,multirow,bm,placeins,xcolor}
\usepackage[hidelinks]{hyperref}
\def\figdir{./BayesianRevenueForecasting_Figures}

\title{\vspace{-1cm} Multivariate Dynamic Modeling for \\ Bayesian Forecasting of Business Revenue}
\author[1]{Anna K. Yanchenko}  
\author[1]{Graham Tierney}
\author[1]{Joseph Lawson}
\author[2]{Christoph Hellmayr}
\author[2]{\\ Andrew Cron}
\author[1]{Mike West}

\affil[1]{Department of Statistical Science,
Duke University, Durham NC 27708-0251. U.S.A.}

\affil[2]{$84.51^\circ$, 100 West 5th Street, Cincinnati, OH 45202.  U.S.A.}
\date{\today}
\begin{document}

\maketitle

\setcounter{page}{0} \thispagestyle{empty}  % title page #=0 & not printed

\begin{abstract}
Forecasting enterprise-wide revenue is critical to many companies and presents several challenges and opportunities for significant business impact.  This case study is based on model developments to address these challenges for forecasting in a large-scale retail company. Focused on multivariate revenue forecasting across collections of supermarkets and product Categories, hierarchical dynamic models are natural: these are able to couple revenue streams in an integrated forecasting model, while allowing conditional decoupling to enable relevant and sensitive analysis together with scalable computation.  Structured models exploit multi-scale modeling to cascade information on  price and promotion activities as predictors relevant across Categories and groups of stores. With a context-relevant focus on forecasting revenue 12 weeks ahead, the study highlights product Categories that benefit from multi-scale information,  defines insights into when, how and why multivariate models improve forecast accuracy, and shows how cross-Category dependencies can relate to promotion decisions in one Category impacting others. Bayesian modeling developments underlying the case study are accessible in custom code for  interested readers. 

\end{abstract}

\newpage

%%%%%%%%%%%%%%%%%%%%%%%%%%%%%%%%%%%%%%%%%%%%%%%%%
\section{Introduction}\label{sec:intro}

Large companies of all kinds define, evolve and rely on enterprise-wise forecasting systems that model and predict many aspects of business development. Central to such business analyses are revenue forecasting components that operate at multiple scales in time and across business enterprises. In large retail supermarket companies, forecasts are impacted by multi-scale influences such as company-wide policy, regional differences, variation across Categories of items bought and sold,  and demand for individual items at individual stores, among many other influences on revenue streams.   In large and diverse supermarket chains, forecast information at multiple levels of aggregation-- devolving to groups of items (Categories) and groups of stores, referred to as Local Store Groups (LSGs) -- are utilized by down-stream decision makers in the enterprise.  In this setting, we discuss aspects of a large case study that evolve modeling approaches to aid and inform these complex decision processes. 

In business sales forecasting, information about demand filters from the bottom-up in terms of consumer behavior that underlies item-level sales. In parallel, information about supply, projected sales targets and macroeconomic considerations filter from the top-down, often in formats that are not easily compatible with statistical forecasting models.  Models generating revenue forecasts for product Categories and groups of stores thus need to integrate bottom-up \textit{and} top-down information.  Forecast outputs also need to be in a form that Category-managers, store-managers and executives can utilize. In major companies with many stores and products, what may appear to be very small improvements in forecast accuracy at the levels of groups of items and groups of  stores can translate to very major revenue impact at the enterprise level; hence modeling developments that yield apparently modest improvements at the \lq\lq micro" levels are of major interest. 
 
In this work, we discuss aspects of a long-term case study of revenue forecasting for a large grocery chain.  There are two primary dimensions of interest: Local Store Groups, groups of policy-similar stores (in terms of geography or management); and Categories, defined groups of similar or related items on sale.  The business setting defines a focus on forecasting revenue 12 weeks ahead for every LSG-Category pair. There are multiple challenges in this and related settings.  While patterns of Category demand are related across LSGs, there is also considerable heterogeneity by LSG and Category.  Sharing information has the potential to improve forecasts, especially for smaller LSGs and Categories, but it is not  obvious at what level to share information due to the heterogeneity.  Key questions arise on how to utilize Category-level information on discounts and pricing, in particular. %  and customer traffic. 
The focus on longer-term forecasting-- a forecast horizon of 12 weeks or more to feed-into longer-term planning and decisions-- defines challenges to all forecasting approaches. 

A number of down-stream business questions are informed by revenue forecasts.  The primary interest is in forecasting for 12 weeks ahead to feed into pricing decisions; even very small improvements in forecast accuracy at LSGs and Category levels can translate to large monetary gains across the system.  The grocery chain is also interested in understanding the roles of pricing and promotion strategies, for both LSGs and Categories, and in exploring ``What-if?'' scenarios where pricing and discounts are altered and the impact of these changes assessed.  This necessitates interpretable models such  that: (i) the roles of such control and predictor variables can be assessed; (ii) users can intervene in the models in informed ways; and (iii) forecast uncertainties are fully characterized for proper use in down-stream decision making.  There is also interest in understanding dependencies between Categories, particularly in relation to possible ``cannibalization'' effects that might occur when one Category is subject to more aggressive discount policies than another that might \lq\lq compete" for customer purchases. There is also the evident need for models to be open, responsive and adaptable over time as realized consumer behavior and grocery demand is inherently time-varying.  
We address these desiderata using customized classes of dynamic linear models~\citep{West-Harrison,PradoFerreiraWest2021} applied to revenue time series at the LSG-Category level, with  
multi-scale extensions~\citep[e.g.][]{BerryWest2018DCMM,West2020Akaike} to represent key aspects of multivariate relationships. 

Statistical forecasting has a long history in revenue management across industries. Models must address basic questions of  seasonality, stochastic variation in demand, price sensitivity, and computational efficiency~\citep[e.g.][]{Weatherford:2016}.  More recently, machine learning and algorithmic approaches have been explored for revenue forecasting. \citet{Pundir:2020} and~\citet{Lei:2021} use random forests and support vector machines, while~\citet{Mishev:2019} and~\citet{Chu:2003} explore deep learning methods.  Such approaches can yield forecast accuracy improvements, especially in short-term forecasting and when time-variation is very limited. They are, however, challenging to interpret and typically neither probabilistic nor dynamic.  Particularly in the retail domain,  Bayesian dynamic models have been successful in terms of forecasting accuracy,  and are substantially preferable in terms of interpretation, openness to intervention, and fully probabilistic forecasting~\citep[e.g.][]{BerryWest2018DCMM, BerryWest2018DBCM, yanchenko2021hierarchical}.  

Our case study also involves methodological contributions. We extend multi-scale models~\citep[e.g.][]{BerryWest2018DCMM, BerryWest2018DBCM, yanchenko2021hierarchical} to allow sharing of discount information,  and represent multivariate structure in pricing %, customer traffic 
and revenue via a recoupled system of univariate models. These are embedded in the case study discussion throughout. 

Section~\ref{sec:data} introduces the retail setting and data. Section~\ref{sec:methods} describes the multi-scale modeling framework, noting the role of the decouple/recouple approach in engendering scalability of multivariate models.  
Section~\ref{sec:results} discusses selected results, highlighting: (i) retail Categories that benefit from multi-scale modeling in improved revenue forecasting, and others that do not; (ii) contexts where forecasts can be improved by joint modeling of pricing, revenue and dependencies across Categories; and (iii) aspects of cross-Category dependencies.  Concluding comments are in Section~\ref{sec:conc}. 

%%%%%%%%%%%%%%%%%%%%%%%%%%%%%%
\section{Setting and Data}\label{sec:data}

The setting is revenue forecasting at the LSG-Category level for a large grocery chain. %, as in~\citet{yanchenko2021hierarchical}.  
 The forecasting level of interest here is across groups of items (Categories) and groups of stores (LSGs).  Each Category is a collection of (a large number of) related items; each LSG is a subset of (a small number of) regionally proximate stores.  LSGs, in general, share traits in terms of discounts offered and pricing, though there is variability across LSGs and Categories.  It is thus important to allow for  variability by LSG and Category, while also allowing information sharing-- as appropriate-- to potentially increase forecast accuracy.   
\begin{figure}[htbp!]
\centering
\begin{subfigure}[b]{0.85\textwidth}
   \includegraphics[width=1\linewidth]{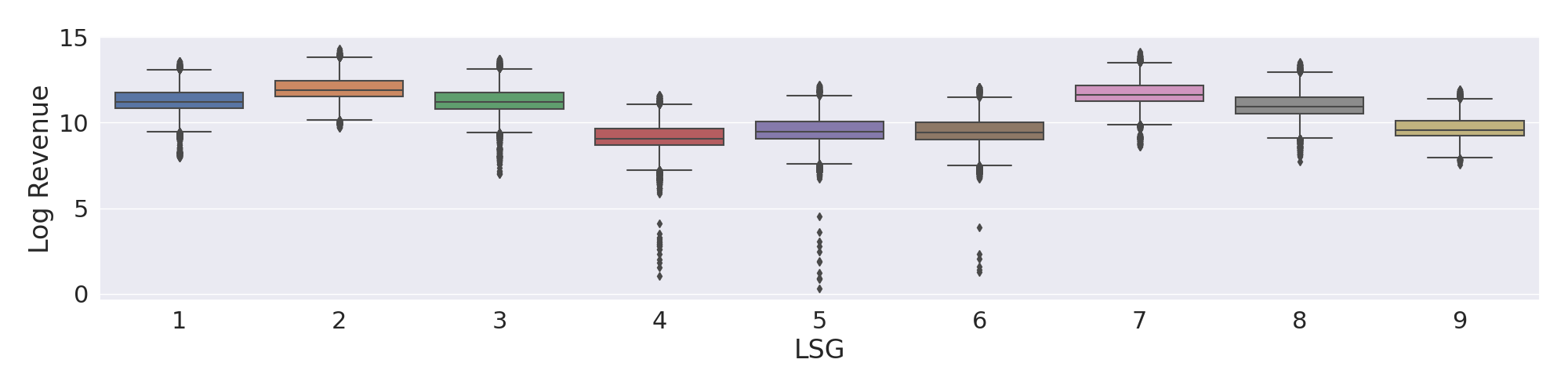}
   \caption{Revenue by Local Store Group.}
\end{subfigure}
\begin{subfigure}[b]{0.85\textwidth}
   \includegraphics[width=1\linewidth]{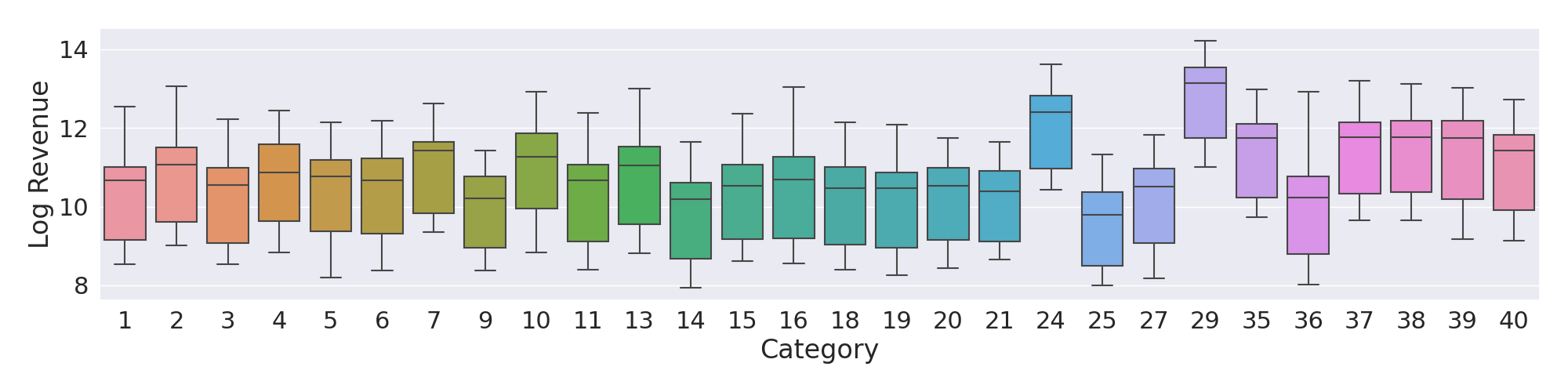}
   \caption{Revenue by Category.}
\end{subfigure}
\caption[Revenue by Local Store Group and Category.]{Log Revenue by (a) Local Store Group and (b) Category over all 104 weeks.  There is variation both by Local Store Group and by Category; a subset of Categories are displayed.}
\label{fig:boxplots-revenue}
%\end{figure}
\bigskip\bigskip
%\begin{figure}[!ht]
%\centering
\begin{subfigure}[b]{0.95\textwidth}
   \includegraphics[width=1\linewidth]{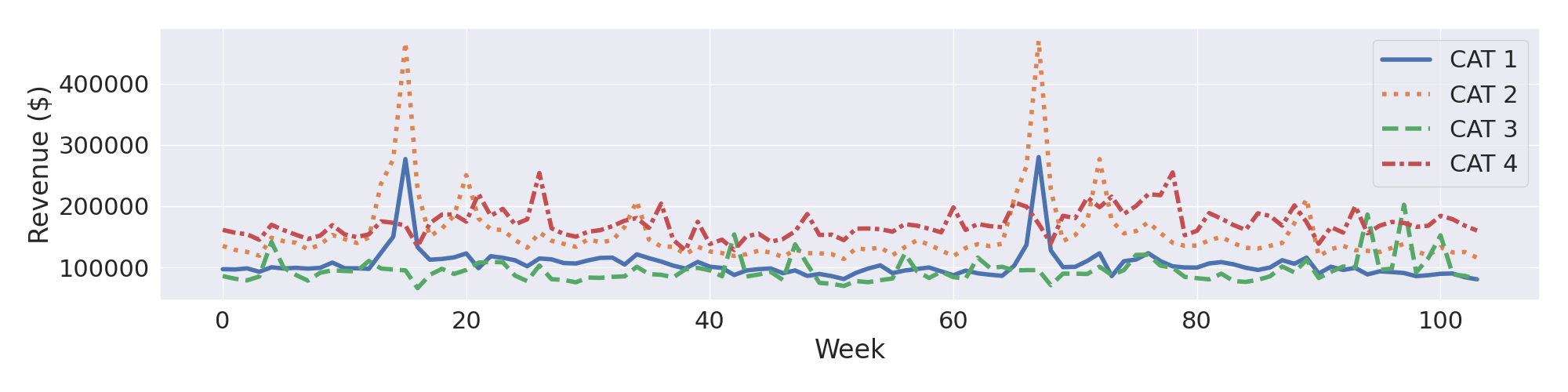}
   \caption{Local Store Group 2.}
\end{subfigure}
\begin{subfigure}[b]{0.95\textwidth}
   \includegraphics[width=1\linewidth]{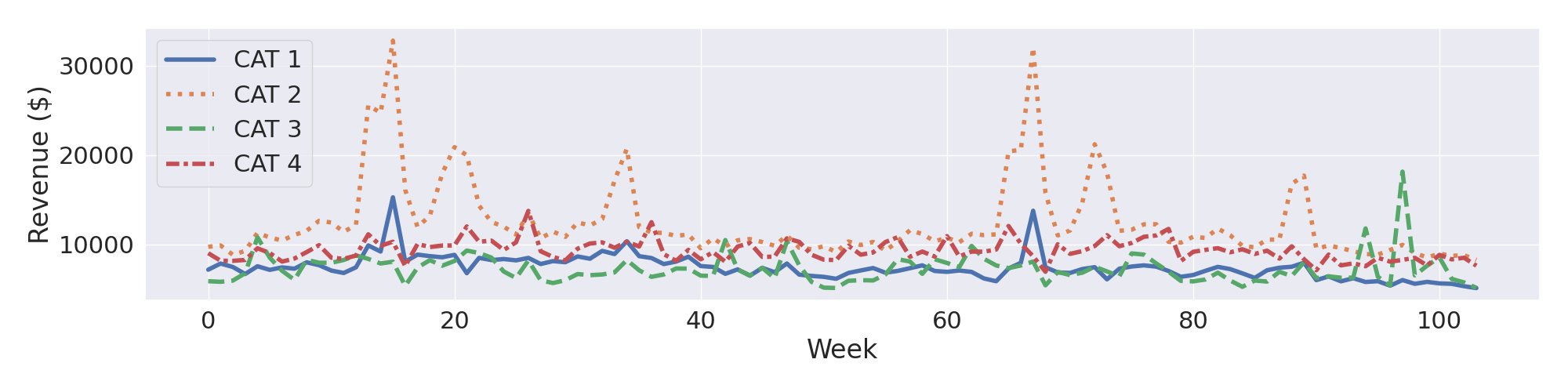}
   \caption{Local Store Group 4.}
\end{subfigure}
\caption[Revenue over time.]{Weekly Revenue for 4 Categories for LSG 2 (large) and LSG 4 (small).  }
\label{fig:lineplots-revenue}
\end{figure}

\begin{figure}[htbp!]
\centering
\begin{subfigure}[b]{0.85\textwidth}
   \includegraphics[width=1\linewidth]{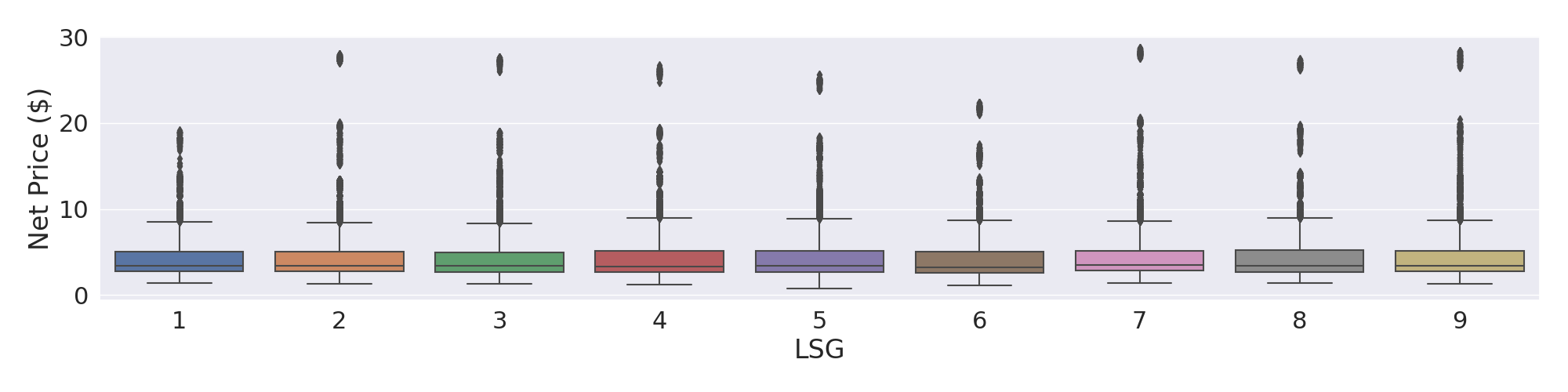}
   \caption{Net Price by Local Store Group.}
\end{subfigure}
\begin{subfigure}[b]{0.85\textwidth}
   \includegraphics[width=1\linewidth]{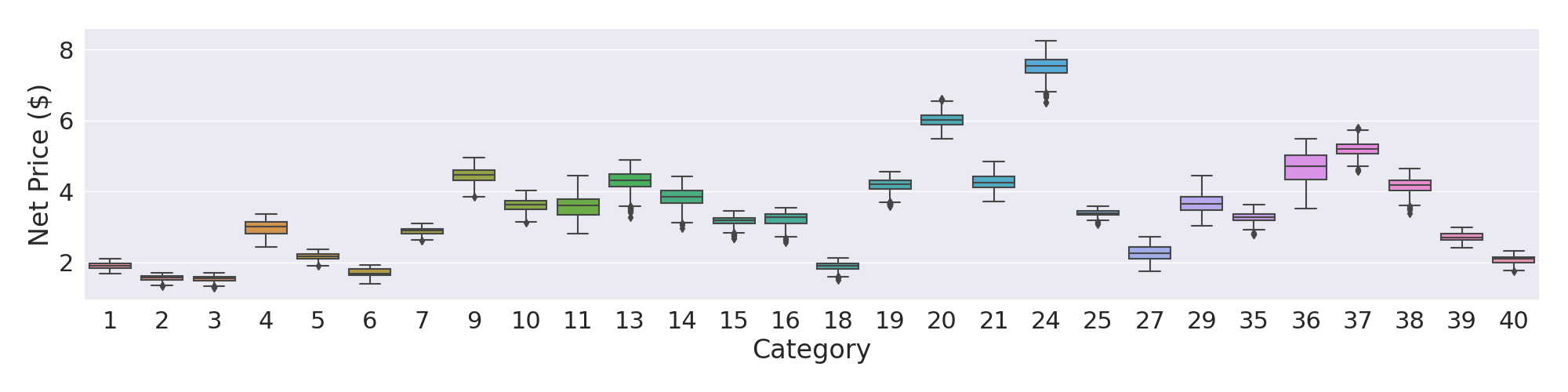}
   \caption{Net Price by Category.}
\end{subfigure}
\caption[Net Price by Local Store Group and Category.]{Net Price by (a) LSG and (b) Category over all 104 weeks.  There is variation by Category, but pricing is very similar across LSGs.}
\label{fig:boxplots-net_prc}
%\end{figure}
\bigskip\bigskip
%\begin{figure}[!ht]
%\centering
\begin{subfigure}[b]{0.9\textwidth}
   \includegraphics[width=1\linewidth]{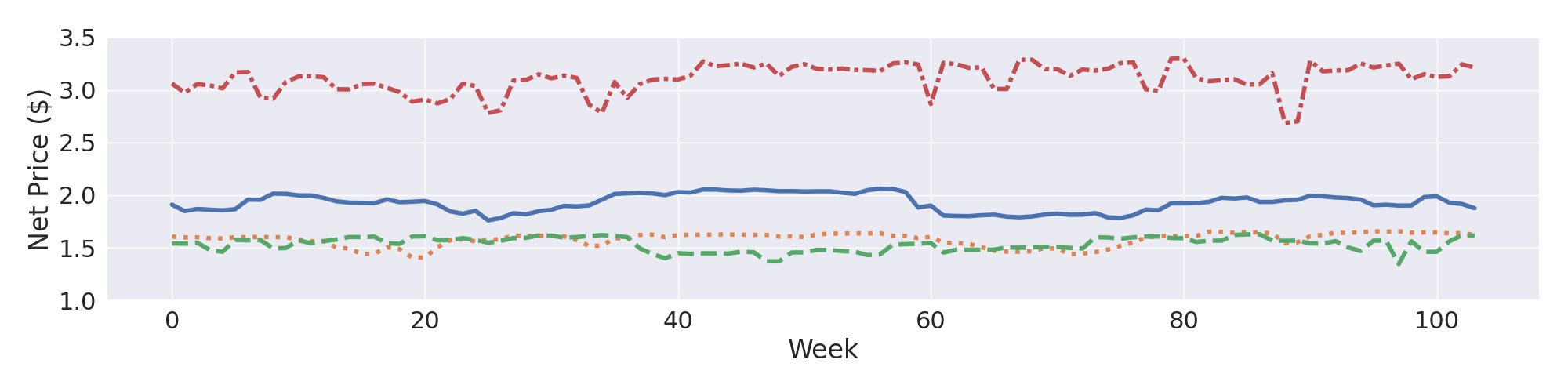}
   \caption{Local Store Group 2.}
\end{subfigure}
\begin{subfigure}[b]{0.9\textwidth}
   \includegraphics[width=1\linewidth]{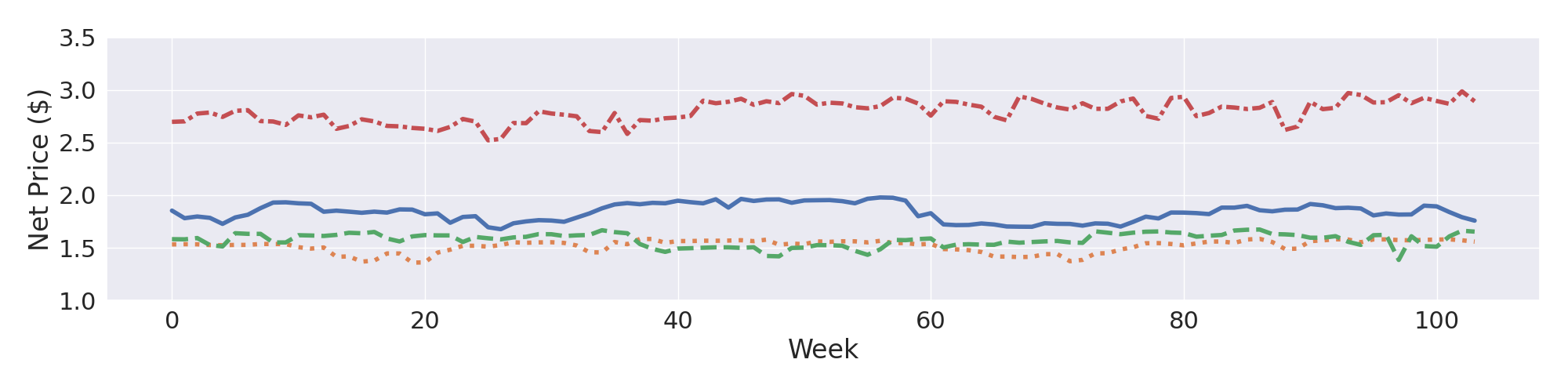}
   \caption{Local Store Group 4.}
\end{subfigure}
\caption[Net Price over time.]{Weekly Net Price for 4 Categories: (a) LSG 2 (large)
and (b) LSG 4 (small). Legend as in \autoref{fig:lineplots-revenue}.}
\label{fig:lineplots-net_prc}
\end{figure}

The data provide 2 calendar years of weekly information for 100 product Categories across 9 LSGs in one geographic region of the USA.  This includes weekly revenue (in \$s) and detailed information about pricing and promotion for each Category and LSG.  Several \lq\lq breadth of discount" measures (weighted averages across items within each Category) exist and we use three: Temporary Price Reduction (TPR) percent, a percent measure of advertising on the front page of leaflets 
(AdFront percent), and a percent measure of special stock displays in the back of stores
(DspBack percent).  Each of these discount measures represents the percentage of items within each Category with each type of discount, weighted by how often each item has historically been purchased. Other information includes the weighted average of discounted price of items within a Category, referred to as the Net Price; this is a quantity that turns out to be quite useful in forecasting weekly LSG-Category level revenue.   Throughout, all revenue results are scaled by a random factor.

%Temporary Price Reduction
%Ad_Front (advertising on front page of leaflet, binary flag)
%Dsp_Back_pct (special stock displays in back of stores, weighted average)

\begin{figure}[htbp!]
\centering
\begin{subfigure}[b]{0.85\textwidth}
   \includegraphics[width=1\linewidth]{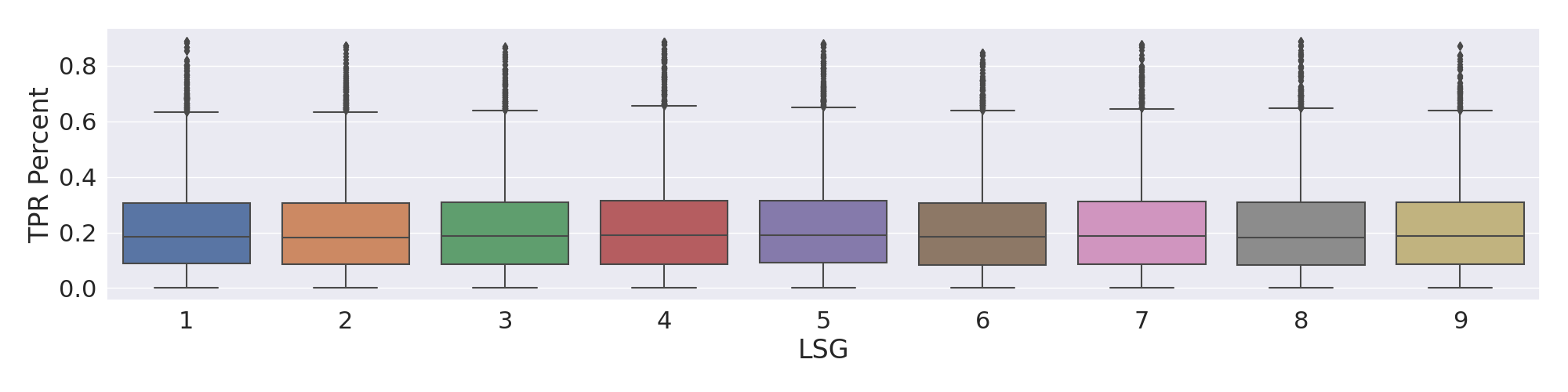}
   \caption{TPR Percent by Local Store Group.}
\end{subfigure}

\begin{subfigure}[b]{0.85\textwidth}
   \includegraphics[width=1\linewidth]{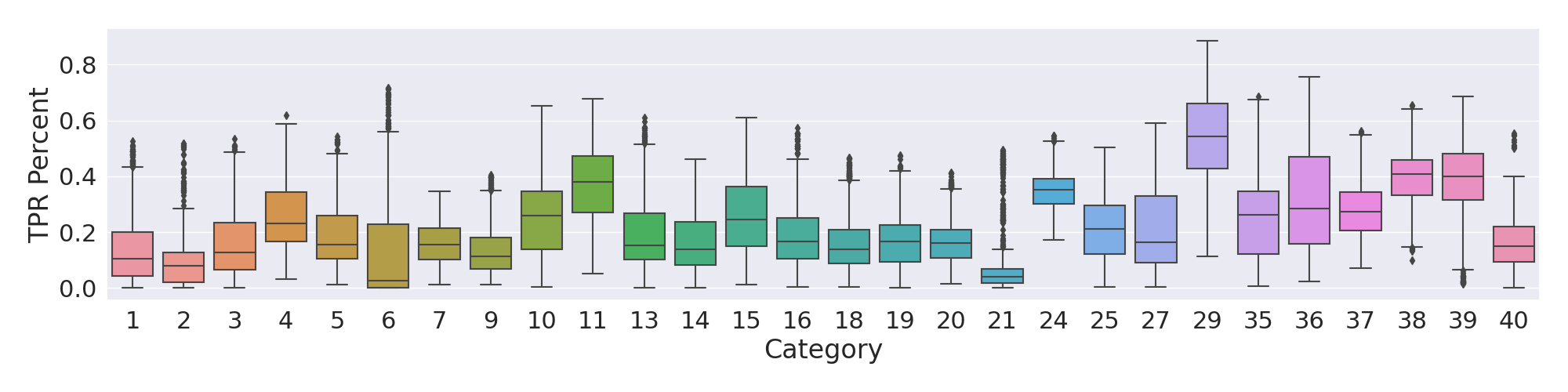}
   \caption{TPR Percent by Category.}
\end{subfigure}
\caption[TPR \%
by Local Store Group and Category.]{TPR Percent by (a) LSG and (b) Category over all 104 weeks.  There is variation by Category, but discounts are very similar across LSGs.}
\label{fig:boxplots-TPR}
%\end{figure}
\bigskip\bigskip
%\begin{figure}[!ht]
%\centering
\begin{subfigure}[b]{0.9\textwidth}
   \includegraphics[width=1\linewidth]{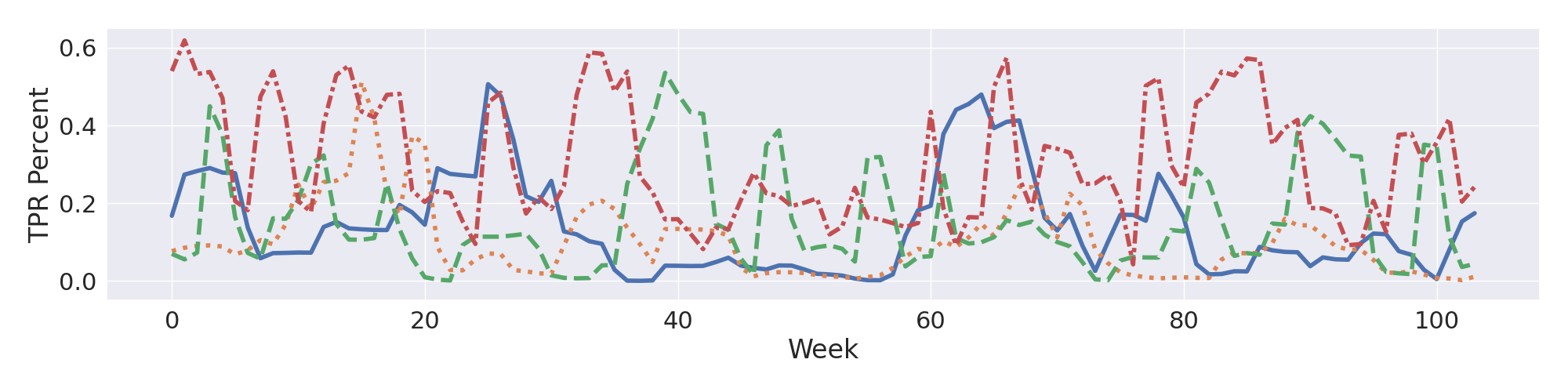}
   \caption{Local Store Group 2.}
\end{subfigure}
\begin{subfigure}[b]{0.9\textwidth}
   \includegraphics[width=1\linewidth]{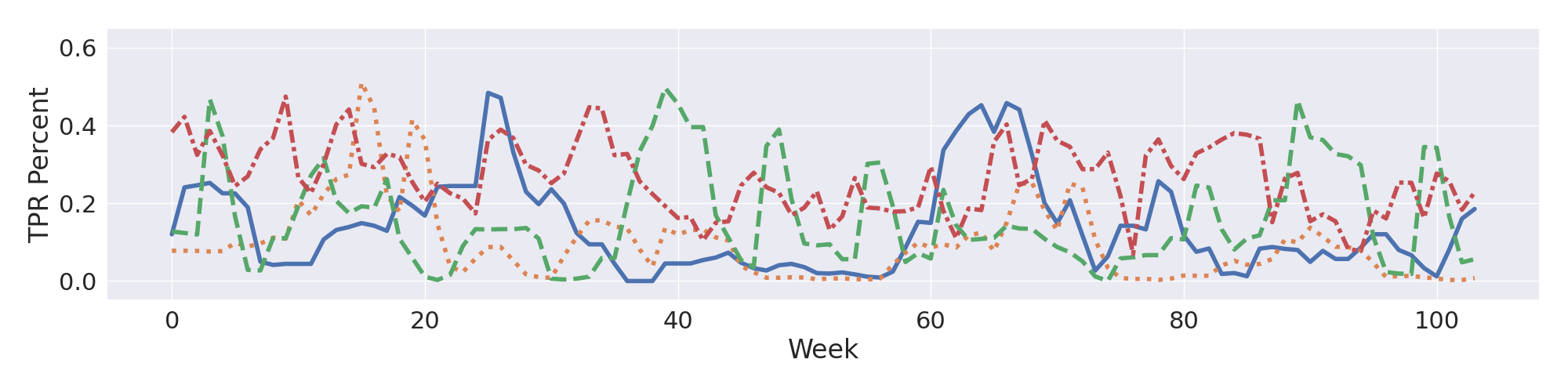}
   \caption{Local Store Group 4.}
\end{subfigure}
\caption[TPR percent over time.]{Weekly TPR\% 
for 4 Categories: (a) LSG 2 (large) and (b) LSG 4 (small). Legend as in \autoref{fig:lineplots-revenue}.}
\label{fig:lineplots-TPR}
\end{figure}

In \autoref{fig:boxplots-revenue}, we see that revenue varies both by LSG and Category.  Over all 104 weeks, however, revenue by Category trends appear similar across LSGs, though different in scale (\autoref{fig:lineplots-revenue}).  While there do appear to be potential holiday effects for some Categories, we do not explicitly take holidays into account here.  Both pricing (\autoref{fig:boxplots-net_prc}) and discounts (\autoref{fig:boxplots-TPR}) tend to be very similar across LSGs, and to vary considerably by Category.  While each LSG has some control over individual discounts for that particular group of stores, there is coordination among the LSGs in terms of pricing and promotion decisions.  Pricing, in particular, tends to be very similar between LSGs over time, and in general, fairly stable for most Categories (\autoref{fig:lineplots-net_prc}).  Variation in the Net Price variable over time and between LSGs is largely a function of discounting, as the Net Price variable is the weighted average of price actually paid by customers after taking any discounts into account.  On the other hand, there is much more variation over time in terms of TPR percent (\autoref{fig:lineplots-TPR}). Again, TPR trends are similar across LSGs, though vary considerably by Category.  TPR percent tends to be the most variable of the three available discount measures.

%%%%%%%%%%%%%%%%%%%%%%%%%%%%%
\section{Methodology}\label{sec:methods}

%We present modeling details including specific methodological extensions of the multi-scale approach relevant to our revenue forecasting setting.

\subsection{Multi-Scale Modeling}\label{subsec:models}
We are interested in forecasting revenue $k=12$ weeks ahead for each LSG-Category pair.  Discount information is set multiple weeks in advance, so discount covariates can be treated as known 12 weeks into the future.  However, Net Price needs to be forecast to be used as a covariate at this forecast horizon, as Net Price depends on the discounts seen by individual customers.  To improve the revenue forecasts at the LSG-Category level, we utilize aggregate multi-scale discount information across LSGs, extending the approach of~\citet{BerryWest2018DCMM}.

Multi-scale analysis enables forecast information from aggregate  levels to inform lower-level forecasts, inherently hierarchical by design.  Multi-scale models are critically interesting alternatives to far more computationally implicated hierarchical  models~\citep[e.g.][]{NIPS2019_8907, sen2019think}.  Multi-scale approaches share information across series while enabling parallel estimation of univariate   models~\citep{BerryWest2018DCMM,Berry:2019,West2020Akaike}.  This enables scaling to large numbers of time series such as are frequently seen in business contexts; computations scale linearly in the number of series.  Importantly, this avoids the need for large, complex Markov chain Monte Carlo  or particle filtering methods, while retaining the ability to improve multi-step ahead forecasts for individual series by incorporating  multi-scale \lq\lq dynamic factor" signals.   Scalability is especially relevant in demand forecasting settings, where there are very many noisy, sparse and heterogeneous individual series.  However, there often exist cross-sectional or other hierarchical structures in this type of data--  across items, for example-- that can be leveraged as aggregate, multi-scale signals to improve forecasts at the lowest level.  Our models here build on this background. 

\begin{figure}[htbp!]
\begin{center}
\includegraphics[width=0.8\linewidth]{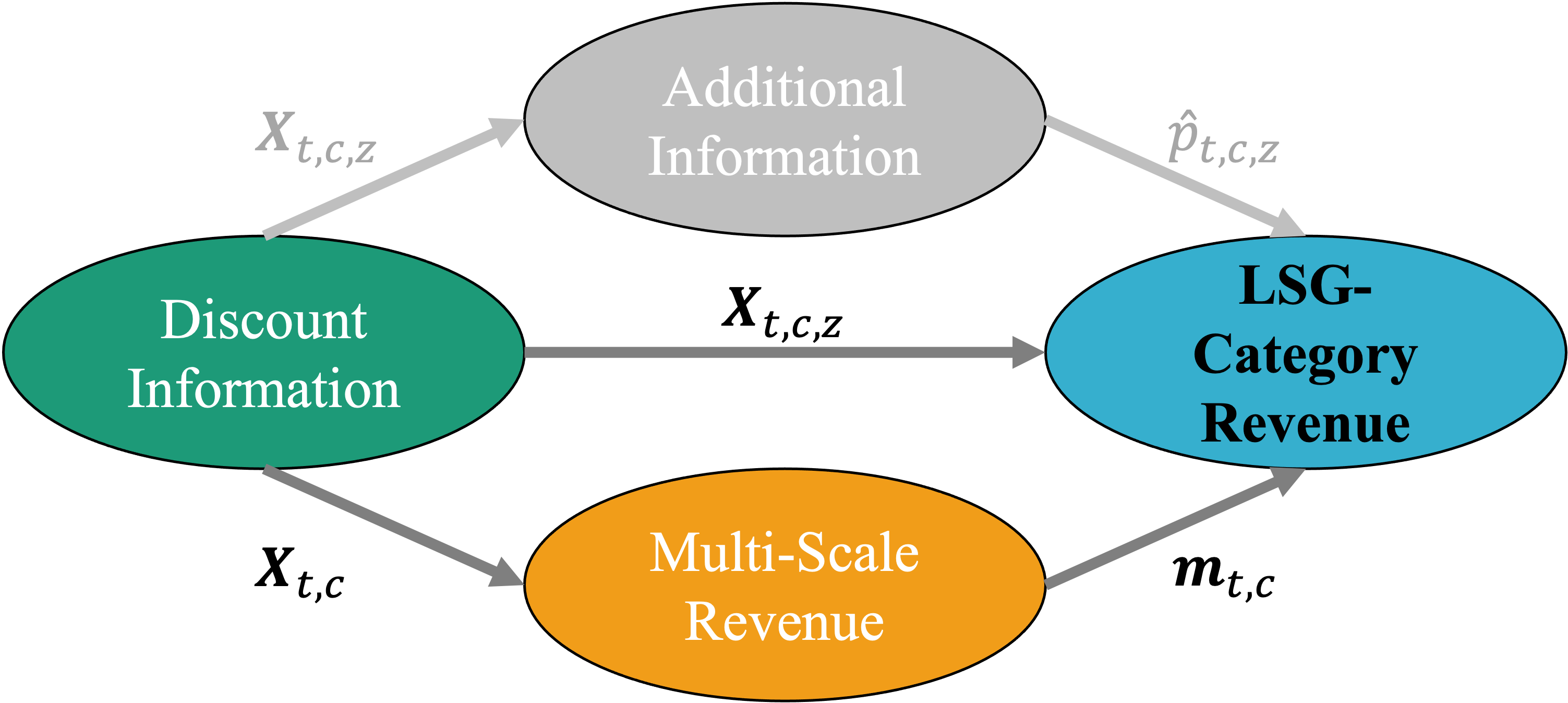}
\vspace*{5mm}
\caption[Baseline multi-scale revenue model diagram.]{Baseline multi-scale revenue modeling at the LSG-Category level.  Information on promotions and multi-scale promotions across LSGs feeds into the models for revenue. }
\label{fig:model}
\end{center}
\end{figure}

Let $Y_{t, c, z}$ be the revenue for week $t$, Category $c$ and LSG $z$ and $Y_{t, c}$ be the revenue aggregated across LSGs for each Category $c$.  Then, let $\bm{X}_{t, c, z}$ be the vector of discount measures (TPR percent, ad front percent and display back percent).  Here  $\bm{X}_{t, c, z}$ is known 12 weeks in advance and we aim to forecast $Y_{t, c, z}$ for all $c,z$ into the future $t$.  Our modeling strategy is to: 

\begin{enumerate}[noitemsep,topsep=0pt]%
\item Model aggregate revenue across LSGs (multi-scale): $Y_{t, c} \vert \bm{X}_{t, c}$.
\item Extract inferred effects of aggregate discounts from model (1): $\bm{m}_{t, c}$.
\item Model revenue: $Y_{t, c, z}\vert \bm{X}_{t, c, z}, \;  \bm{m}_{t, c}$.
\end{enumerate}
This model
for revenue depends on LSG-Category specific discount information ($\bm{X}_{t, c, z}$) and multi-scale discount information across LSGs ($\bm{m}_{t, c}$); see~\autoref{fig:model}.  This defines a flexible baseline model. 
Section~\ref{subsec:improve} discusses extensions to include Category pricing information 
%and customer traffic information 
that can yield revenue forecasting improvements.

This hierarchical, multi-scale approach allows each LSG-Category pair to ``see'' common, aggregate revenue responses to discounts differently and allows for sharing of information and personalization of the common trends for each specific LSG.  This approach increases  forecast accuracy for many LSG-Category pairs for 12-week ahead revenue forecasts, in particular for smaller LSGs that build on information from larger LSGs. On a key technical point, we use  \lq\lq plug-in" point forecasts $\bm{m}_{t,c}$ of the multi-scale effects of discount predictors, choosing the current (time $t$) posterior mean of the effect in the aggregate model.  This  under-states uncertainty in resulting revenue forecast distributions as it ignores uncertainty about aggregate discount effects. Applied evaluations lead us to accept this practical side-step of full uncertainty characterization, as it has  modest practical impact. At the costs of more extensive computation it is, of course, easy to extend the analysis to  include full uncertainty characterization, repeating the analysis with Monte Carlo samples of the discount effect; see~\citet{BerryWest2018DCMM} in related models. This more computationally intensive analysis, across numerous LSGs and Categories, can aid in understanding how relevant or-- in  this case study-- practically limited, is the impact of this second-order uncertainty analysis.

\subsection{Dynamic Linear Models}\label{subsec:DLMs}

DLMs define the core class of time series models for all levels in the multi-scale setting of~\autoref{fig:model}. For a generic univariate time series $y_t$ observed at discrete times $t = 1, \ldots, T$, information at time $t$ is denoted by $\mathcal{D}_t = \{y_t, \mathcal{D}_{t-1}, \mathcal{I}_{t-1}\}$ where $\mathcal{I}_{t-1}$ represents any additional relevant information beyond the observed data.  A DLM has the form
\begin{equation}\label{eq:dlm}
\begin{split}
y_t &= \bm F_t'\bm\theta_t + \nu_t, \enspace \nu_t \sim \mathcal{N}(0, v_t),\\
\bm\theta_t &= \bm G_t\bm\theta_{t-1} + \bm\omega_t, \enspace \enspace \bm\omega_t\sim \mathcal{N}(\bm 0, \bm W_t),
\end{split}
\end{equation}
where:
\begin{itemize}[noitemsep,topsep=0pt]%
 \item $\bm F_t$ is a matrix of known covariates at time $t$,
 \item $\bm \theta_t$ is the state vector, which evolves via a first-order Markov process,
 \item $\bm G_t$ is a known state evolution matrix, 
 \item $\bm \omega_t$ is the stochastic innovation vector, with the $\bm \omega_t$ independent over time, and 
 \item $\bm W_t$ is the known innovation variance matrix at time $t.$
\end{itemize}
Sequential learning in the DLM proceeds naturally via computationally easy updates and forecasting algorithms. Analysis at the level of each univariate series is standard~\citep{West-Harrison,PradoFerreiraWest2021}.
%proceeds as follows for the time $t-1$ evolve-predict-update cycle %(following~\citealp{Berry:2019}):
%\begin{enumerate}[noitemsep,topsep=0pt]%
%\item Posterior at $t-1$: $\left(\bm \theta_{t-1} \vert \mathcal %D_{t-1}, \mathcal I_{t-1}\right) \sim \mathcal{N}\left(\bm m_{t-1}, \bm %C_{t-1}\right).$
%\item Prior at $t$: $\left(\bm \theta_{t} \vert \mathcal D_{t-1}, %\mathcal I_{t-1}\right) \sim \mathcal{N}\left(\bm a_t, \bm R_t\right)$ %%with $\bm a_t = \bm G_t\bm m_{t-1}$ and $\bm R_t = \bm G_t\bm C_{t-1} %\bm G_t' + \bm W_t$.
%\item Forecast $y_t$ 1-step ahead: $p(y_t \vert \mathcal D_{t-1}, %\mathcal I_{t-1}) = \mathcal{N}(f_t, q_t)$ with $f_t = \bm F_t'\bm %a_t$, $q_t = \bm F_t'\bm R_t\bm F_t + v_t$.
%\item Posterior at time $t$: $\left(\bm \theta_t | \mathcal %D_t\right)\sim\mathcal{N}\left(\bm m_t, \bm C_t\right)$ given by
%$$\bm m_t = \bm a_t + \bm R_t\bm F_t (y_t - %f_t)/q_t\enspace\mbox{and}\enspace \bm C_t = \bm R_t - \bm R_t\bm %F_t\bm F_t'\bm R_t' /q_t.$$
%\end{enumerate}
%\noindent This completes the time $t-1$ to $t$ evolve-predict-update cycle.
%DLMs are probabilistic, interpretable and computationally efficient and %allow us to explore the various down-stream questions outlined in %Section~\ref{sec:intro}.  Additionally, while we fit univariate DLMs to %each LSG-Category pair, in Section~\ref{subsec:cross-cat}, we %can use them to explore cross-Category questions.  

\subsection{Modeling Details} 

Revenue is modeled on the log scale using normal DLMs with a trend term and additional covariates;  each univariate DLM has the $\bm F_t$ vector with a leading element of 1 followed by entries representing potential seasonal components and known predictor/covariate values. Among the latter,  the aggregate revenue model for $Y_{t, c}$ uses the average discounts across LSGs, $\bm{X}_{t, c}$, as additional covariates and has yearly seasonality represented by the fundamental (52 week) harmonic model component.  The LSG-Category revenue model for $Y_{t, c, z}$, has  multi-scale discount information included as predictor values; here $\bm F_{t,c,z}$ has elements 
%$\bm{X}_{t, c, z}\odot \bm{m}_{t, c} = 
$X^{TPR}_{t, c, z}m^{TPR}_{t,c},$
$X^{Ad Front}_{t, c, z}m^{Ad Front}_{t,c}$ and 
$X^{Dsp Back}_{t, c, z}m^{Dsp Back}_{t,c}$ 
as covariates, again with yearly seasonality defined by the first harmonic.  All models use the same specific state evolution discount factors to define rates of change over time of state vectors. This completes the basic DLM outlook for each univariate revenue series.

In terms of customized predictor information,  Category price discount covariates that are negligible over all weeks are not included  (some Categories are rarely discounted, especially various alcohol Categories).  Similarly, covariates that  are static for many weeks have some small amount of noise added to them to stabilize the modeling; this is a common approach in machine learning and has connections to ridge regression.  Here, we add noise to control variables to (1) stabilize inference when there is not much variation in the covariates, and (2) to reflect potential noise in the estimation of these control variables out to 12 weeks in advance, for some increased robustness in the models for practical application. All LSG-Categories pairs are modeled separately as univariate DLMs as described in Section~\ref{subsec:DLMs}. Recoupling is then induced by sharing information within the over-arching multi-scale framework. Analysis is implemented in PyBats~\citep{PyBats}.

%%%%%%%%%%%%%%%%%%%%%%%%%%
\section{Selected Results}\label{sec:results}

Models were fit and evaluated over the first year of data to define selection of DLM discount factors. The detailed forecasting analysis and selected evaluations are based on then running the analyses sequentially over the second year of data with out-of-sample forecasts generated each week for the following 12 weeks.  Empirical forecast accuracy measures are all on the 12-week horizon.  Section~\ref{subsec:results-ms} gives selected examples where multi-scale modeling improves revenue forecasts and others where it does not.  Section~\ref{subsec:improve} highlights situations where adding  information to the multi-scale models is shown to improve revenue forecasting, with rationalization and discussion of business implications.  Section~\ref{subsec:cross-cat} explores aspects of dependencies across Categories with a view to advising potential competing goals in Category-wide pricing and discount strategies. Throughout, all revenue results are scaled by a random factor.

%Throughout the results, our main metric of interest is mean absolute percent error, or MAPE. % MAPE is widely used in business contexts and allows for easy comparisons across settings and %contexts, as it is on the percent scale.  For forecasts $f_{1:T}$ and observed values %$y_{1:T}$ the MAPE loss functions is defined as:
%\begin{equation}\label{eq:APE}
%\mathcal{L}_{MAPE}(y,f) = \dfrac{1}{T}\sum_{t=1}^T \biggl| 1 - \dfrac{f_t}{y_t}\biggl|.
%\end{equation}
%Under the MAPE loss function, the optimal point forecast $f_t$ is the (-1)-median defined as %the median of the pdf $g(y_t) \; \propto \; p(y_t)/y_t$, where $p(y_t)$ is the forecast %probability density function. 

\subsection{Multi-Scale Revenue Forecasting}\label{subsec:results-ms}

\subsubsection{Some Aggregate Results}

\begin{figure}[hb!]
\begin{center}
\includegraphics[width=0.5\linewidth]{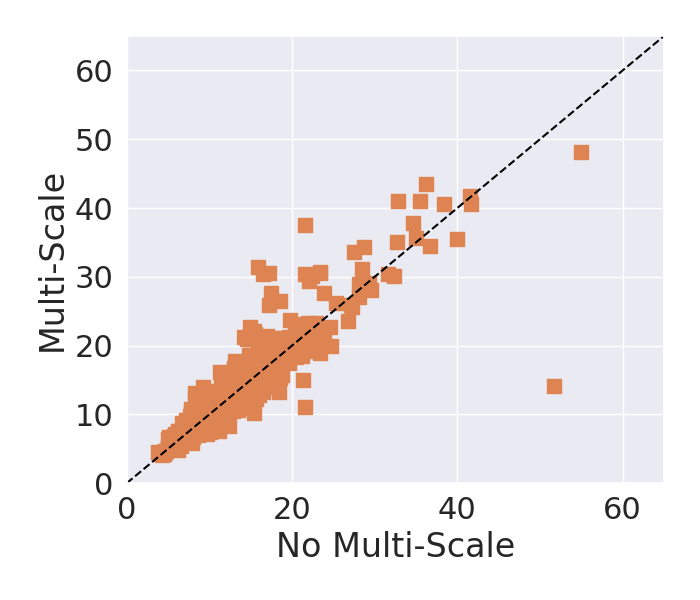}
 %\vspace*{2mm}
\caption{Empirical MAPE comparisons of multi-scale model versus baseline model. Each point represents the one-year average of 12-week ahead forecast MAPE for one LSG-Category pair. }
\label{fig:scatter-compare}
\end{center}
\end{figure}

A first interest is in identifying Categories and LSGs where there are forecast improvements using the multi-scale analysis that shares discount information across LSGs, as described in Section~\ref{sec:methods}.  Using the MAPE metric, the results vary by LSG-Category pair, as seen in~\autoref{fig:scatter-compare}.  About 45\% of the LSG-Category pairs benefit from the inclusion of multi-scale discount information, having lower MAPE values. Again, at this enterprise-wide level of forecasting, even small very improvements in MAPE can lead to large increases in revenue, so these cases are of key interest. Then, identifying cases that are better forecast without the multi-scale information is just as important; these LSG-Category pairs will be forecast using their individual models.  

%Median and standard deviation MAPE values across all LSG-Category pairs are given in ~\autoref{tab:MAPE-models}.  
%\begin{table}[htp]
%\caption{Comparison of the MAPE median and standard deviation values across LSG-Category pairs %for three revenue models.  The first model, ``No Multi-Scale'', uses only discount information at the %LSG-Category level.  The ``Multi-Scale'' model includes multi-scale discount information %across LSGs, as described in Section~\ref{sec:methods}.}
 %\vspace*{5mm}
%\begin{center}
%\begin{tabular}{c|c}
%Model & MAPE \\\hline
%No Multi-Scale & $10.54\pm 6.20$ \\
%Multi-Scale & $10.61\pm 6.05$ \\
%\end{tabular}
%\end{center}
%\label{tab:MAPE-models}
%\end{table}%

\subsubsection{Revenue Forecasts}

We now focus on specific LSG-Category examples that benefit from multi-scale information.   In addition to the forecasts themselves, we look at the regression effect of the discount information from the multi-scale model to illuminate the impact of the multi-scale information.  For each Monte Carlo sample  $\bm{\theta}^{(i)}_{t, c}$ of the state vector from the multi-scale model across LSGs, the discount regression effect $\tilde{m}_{t, c}^{(i)}$ is
$$\tilde{m}_{t, c}^{(i)}= X_{t, c}^{TPR}\theta^{TPR, (i)}_{t, c} + X_{t, c}^{Ad Front}\theta^{Ad Front, (i)}_{t, c} + X_{t, c}^{Dsp Back}\theta^{Dsp Back, (i)}_{t, c}.$$ 
This represents the overall impact of the multi-scale discount information. 

Some general points and findings are noted first.  Forecasting 12 weeks ahead is challenging. An evaluation on 1 week ahead forecasts could be misleading in terms of the main longer-term horizon of interest.  Then, we find that in the cases where multi-scale information improves the forecasts at the 12-week horizon, it also does at the 1 week ahead forecast horizon.  Further, multi-scale information can improve the forecasts of both large and small LSGs.  Additionally, Category discount information is absolutely critical to include in the revenue forecasting models, either as multi-scale information or not.  As the main control variable, the discount information is able to produce good forecasts alone for the majority of LSGs and Categories.  Finally, some Categories have clear and strong holiday effects.  With only two years of data here, there is not enough information to estimate holiday effects directly, but we discuss possible approaches to addressing holiday information in more detail in Section~\ref{subsec:improve}.

\begin{figure}[hp!]
\centering
\begin{subfigure}[b]{0.95\textwidth}
   \includegraphics[width=1\linewidth]{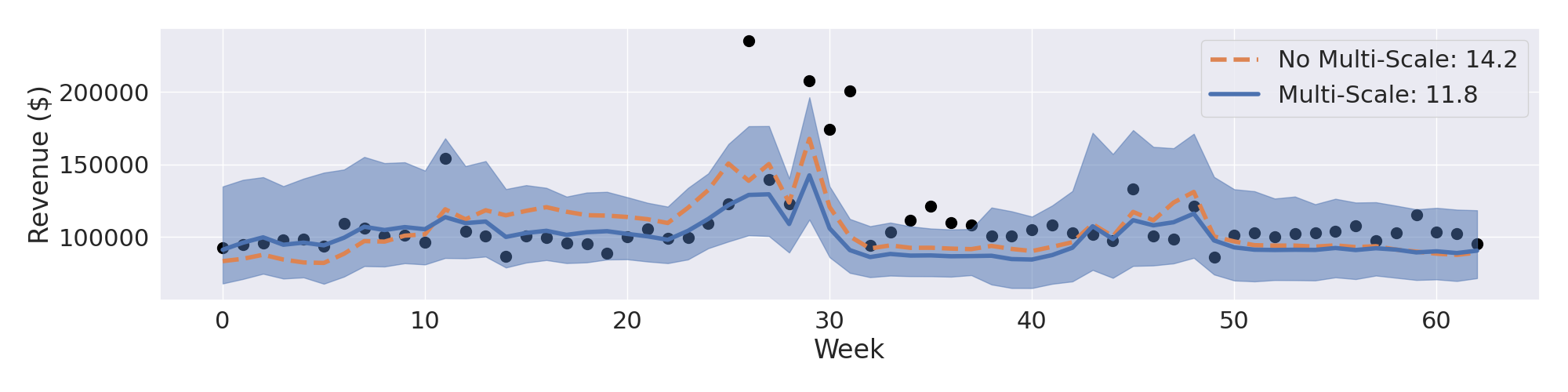}
   \caption{Forecasts - Local Store Group 2.}
\end{subfigure}
\begin{subfigure}[b]{0.95\textwidth}
   \includegraphics[width=1\linewidth]{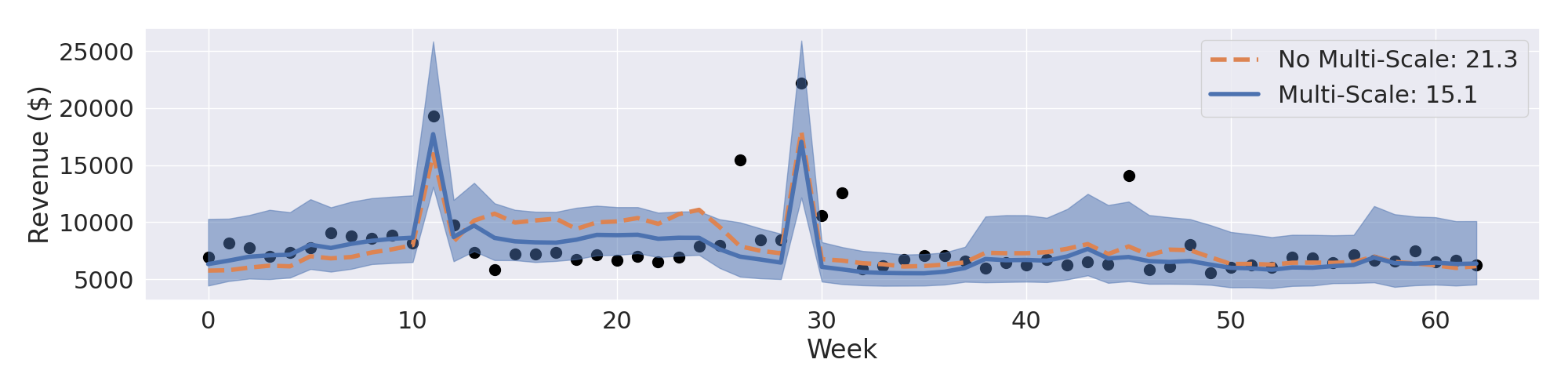}
   \caption{Forecasts - Local Store Group 4.}
\end{subfigure}
\begin{subfigure}[b]{0.95\textwidth}
   \includegraphics[width=1\linewidth]{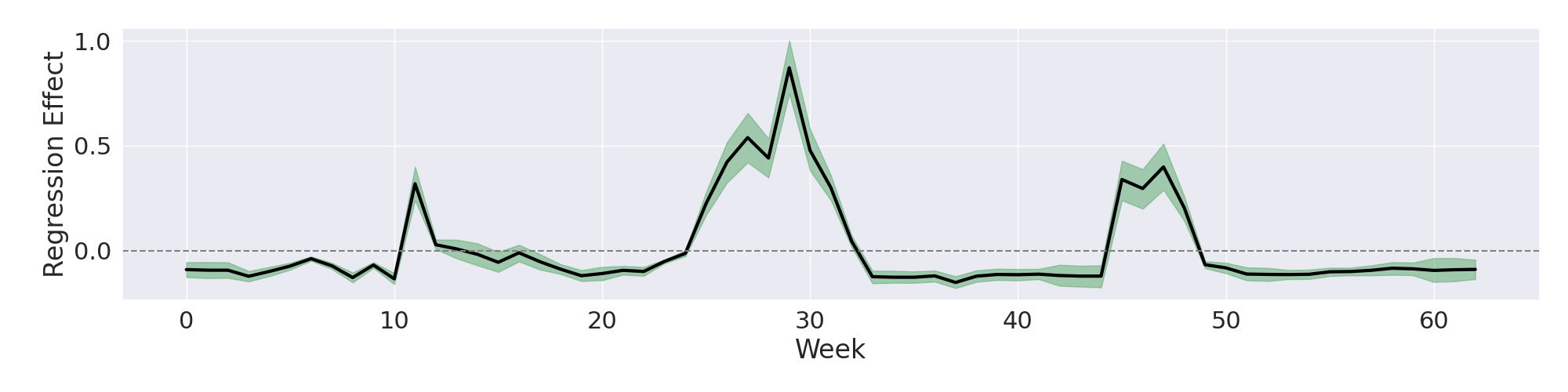}
   \caption{Regression Effect.}\label{fig:sugars-reg}
\end{subfigure}
\caption{Frames (a) and (b) show 12-week ahead forecasts from the multi-scale %(in blue and solid lines) 
and the no multi-scale models %(in orange and dashed lines) 
for the Sugar \& Sweeteners Category for two LSGs. Average MAPE values over the year are shown in the legends.  The point forecasts are MAPE optimal, shading shows 90\% credible intervals in the multi-scale model, and points are the observed revenue values.  For all LSGs, the inclusion of multi-scale information improves the forecasts. Frame (c) shows the on-line estimated regression effects, with 90\% credible intervals, of the combined  discount predictor information.}
\label{fig:sugars-main}
\end{figure}

One Category that particularly benefits from multi-scale information is the Sugars \& Sweeteners Category, with forecasts for two LSGs and the multi-scale regression effect shown in \autoref{fig:sugars-main}.  Across both larger and smaller LSGs, the inclusion of multi-scale discount information defines MAPE optimal forecasts  
that are more accurate than those from the no multi-scale model, especially over weeks 10 and 30.    For Sugars \& Sweeteners, around weeks 10-20 in \autoref{fig:sugars-reg}  there is a dynamic, negative discount regression effect, compared to the rest of the weeks; this translates to lower forecasts from the multi-scale model compared to the no multi-scale model in \autoref{fig:sugars-main}.  This negative regression effect pulls the forecasts down in this region, leading to more accurate forecasts.  This response to discounts in terms of the revenue is shared across LSGs and well captured by the multi-scale model, leading to improved forecasts for this specific Category.  Additionally, in both the forecasts and regression effect in \autoref{fig:sugars-main}, there are strong holiday effects around week 30 (the week of December 15).   

\begin{figure}[!htp]
\centering
\begin{subfigure}[b]{0.95\textwidth}
   \includegraphics[width=1\linewidth]{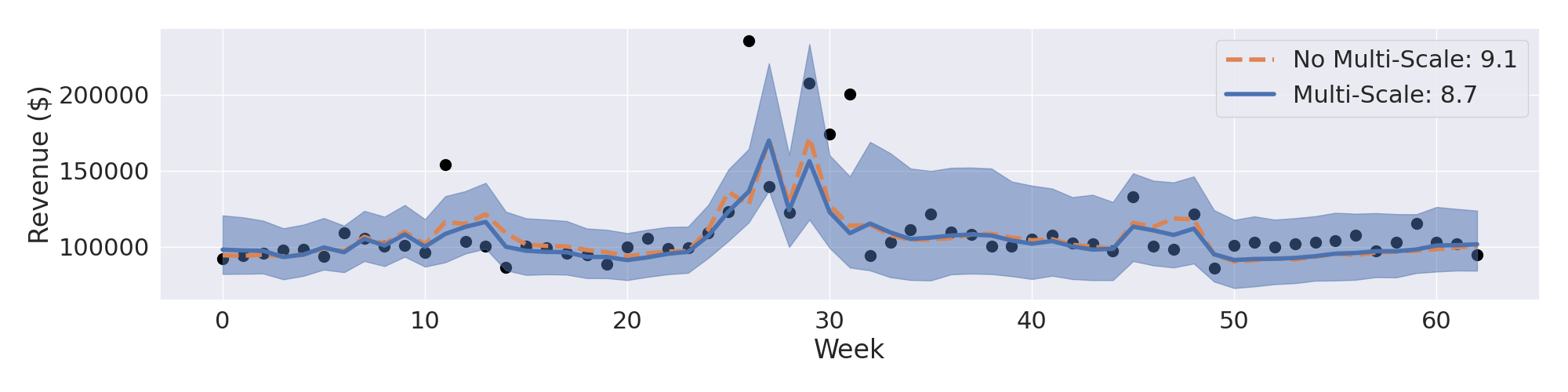}
   \caption{Forecasts - Local Store Group 2.}
\end{subfigure}
\begin{subfigure}[b]{0.95\textwidth}
   \includegraphics[width=1\linewidth]{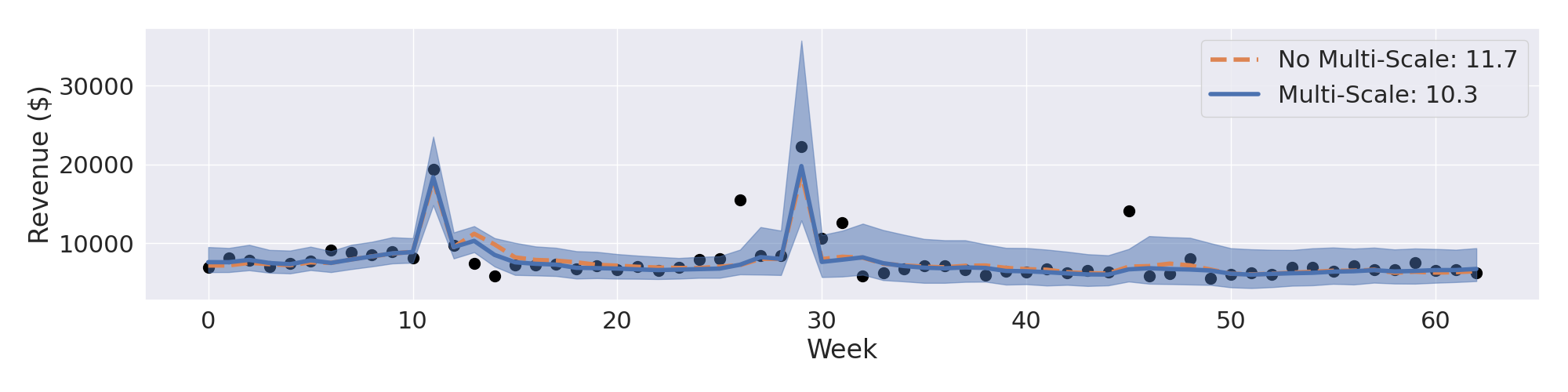}
   \caption{Forecasts - Local Store Group 4.}
\end{subfigure}
\caption{Frames (a) and (b) show 1 week ahead forecasts for the Sugar \& Sweeteners Category in the same two LSGs, and in the same format, as in \autoref{fig:sugars-main}.}
\label{fig:sugars-k1}
\end{figure}

\autoref{fig:sugars-k1} shows similar forecast summaries at the 1 week ahead  horizon. While overall forecast accuracy is naturally higher than that for the 12-week horizon, note that the multi-scale model still leads to improved forecasts for these LSG-Category pairs.  

\FloatBarrier\newpage

\begin{figure}[hp!]
\centering
\begin{subfigure}[b]{0.95\textwidth}
   \includegraphics[width=1\linewidth]{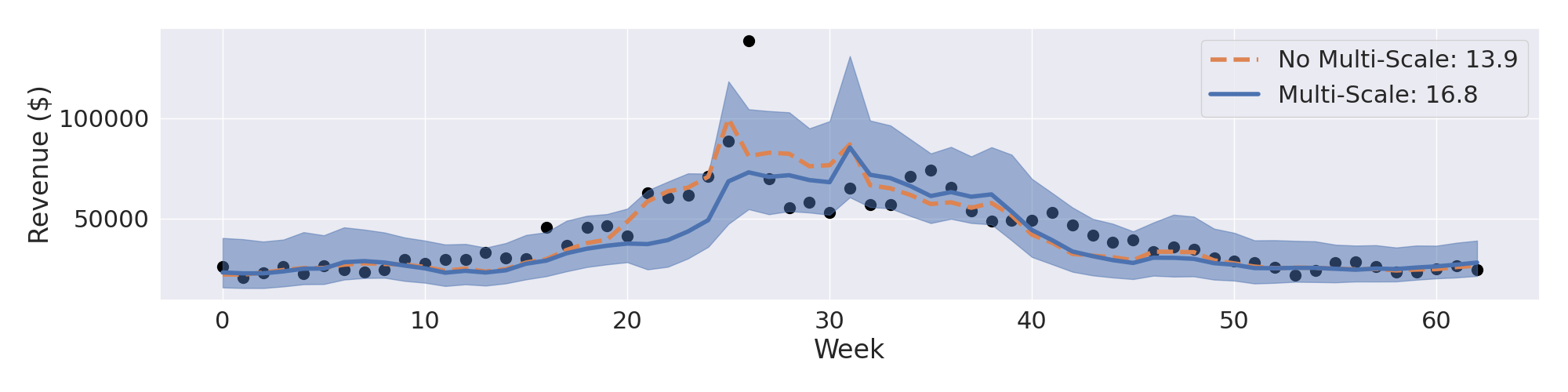}
   \caption{Forecasts - Local Store Group 1.}
\end{subfigure}
\begin{subfigure}[b]{0.95\textwidth}
   \includegraphics[width=1\linewidth]{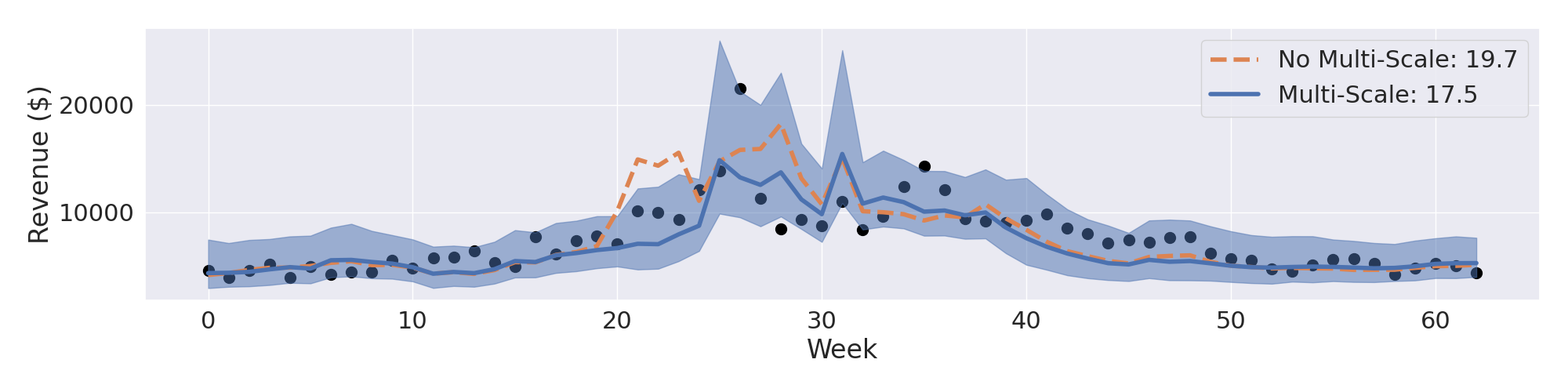}
   \caption{Forecasts - Local Store Group 6.}
\end{subfigure}
\begin{subfigure}[b]{0.95\textwidth}
   \includegraphics[width=1\linewidth]{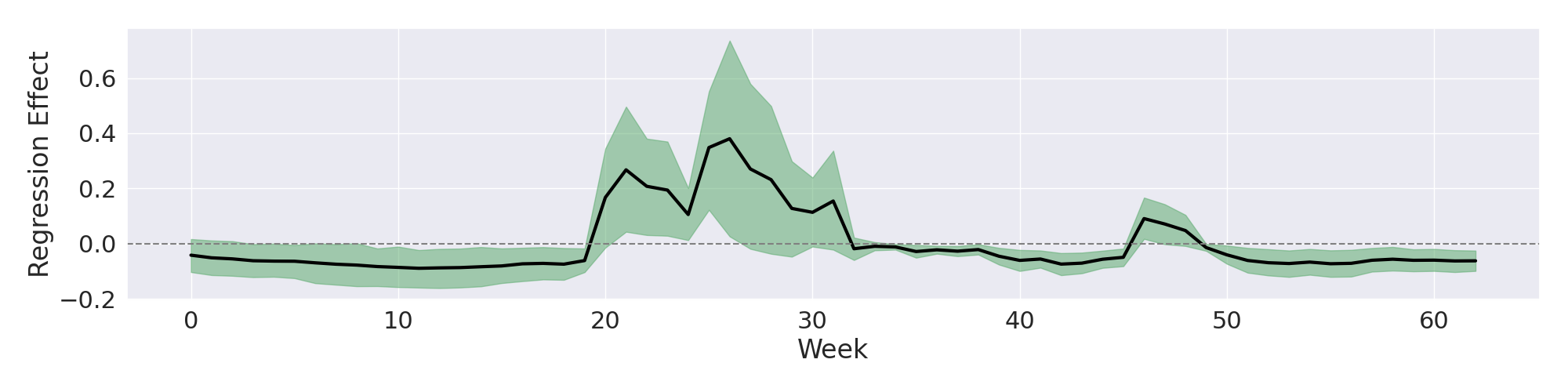}
   \caption{Regression Effect.}\label{fig:broth-reg}
\end{subfigure}
\caption{Summary 12-week ahead forecast and regression effect graphs for the Broth/Dry Soup Category for two LSGs. Details and format  as in \autoref{fig:sugars-main}.}
\label{fig:broth-main}
\end{figure}

\begin{figure}[!htp]
\centering
\begin{subfigure}[b]{0.95\textwidth}
   \includegraphics[width=1\linewidth]{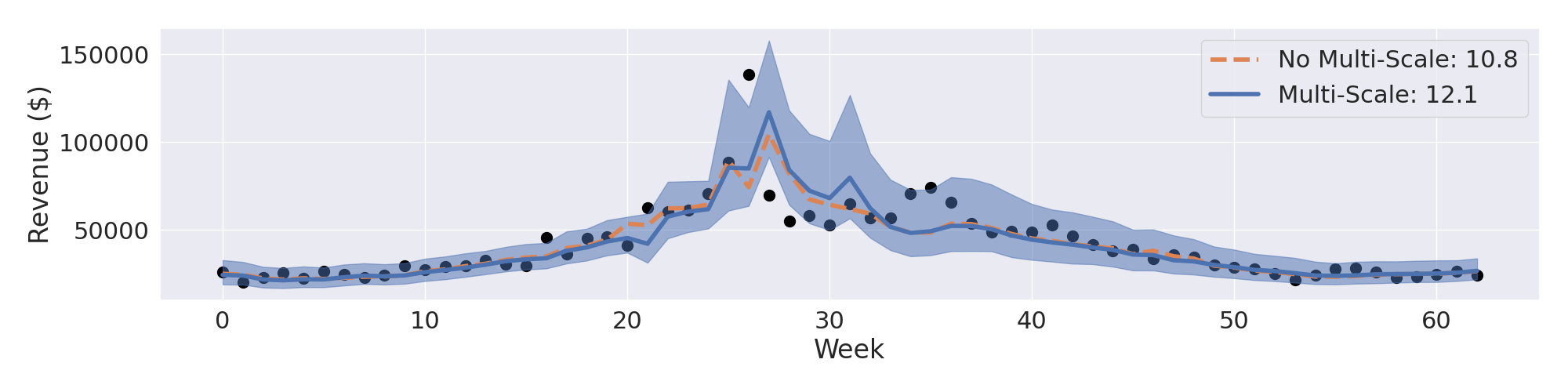}
   \caption{Forecasts - Local Store Group 1.}
\end{subfigure}
\begin{subfigure}[b]{0.95\textwidth}
   \includegraphics[width=1\linewidth]{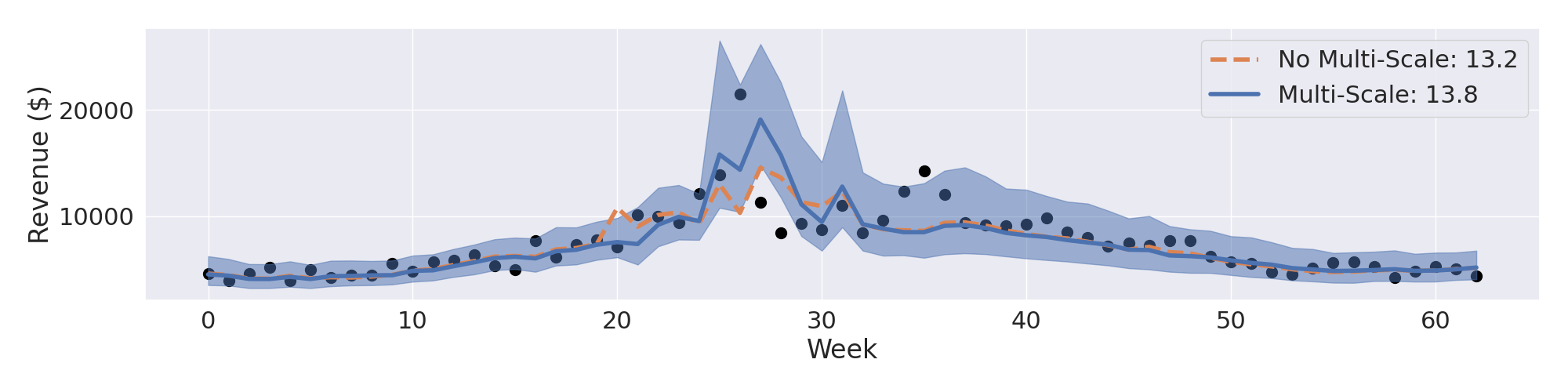}
   \caption{Forecasts - Local Store Group 6.}
\end{subfigure}
\caption{Summary 1 week ahead forecast graphs for the Broth/Dry Soup Category for two LSGs. Details and format as in \autoref{fig:sugars-k1}.}
\label{fig:broth-k1}
\end{figure}

Broth/Dry Soup is an example of a Category where the value of the multi-scale information varies by LSG.  In the larger LSGs in \autoref{fig:broth-main}, there is little benefit from the multi-scale information and the multi-scale model tends to under-forecast around weeks 20-30.  However, there is real benefit from the multi-scale information 
for the smaller LSG. This is a common finding in hierarchical models: smaller groups (here LSGs) can benefit more from sharing of information across larger groups due to the increased shrinkage on smaller groups.  The multi-scale discount information improves forecasts the most for smaller LSGs generally.  In this example, note also the change in regression effect around weeks 20-30, shown in \autoref{fig:broth-reg}. The multi-scale regression effect tends to lead to better forecasts for this time period for the smaller LSGs, as compared to the larger LSGs which under-forecast here. Forecasts for both models also naturally improve at the shorter, 1 week ahead, forecast horizon; see \autoref{fig:broth-k1}.

\FloatBarrier\newpage

\begin{figure}[hp!]
\centering
\begin{subfigure}[b]{0.95\textwidth}
   \includegraphics[width=1\linewidth]{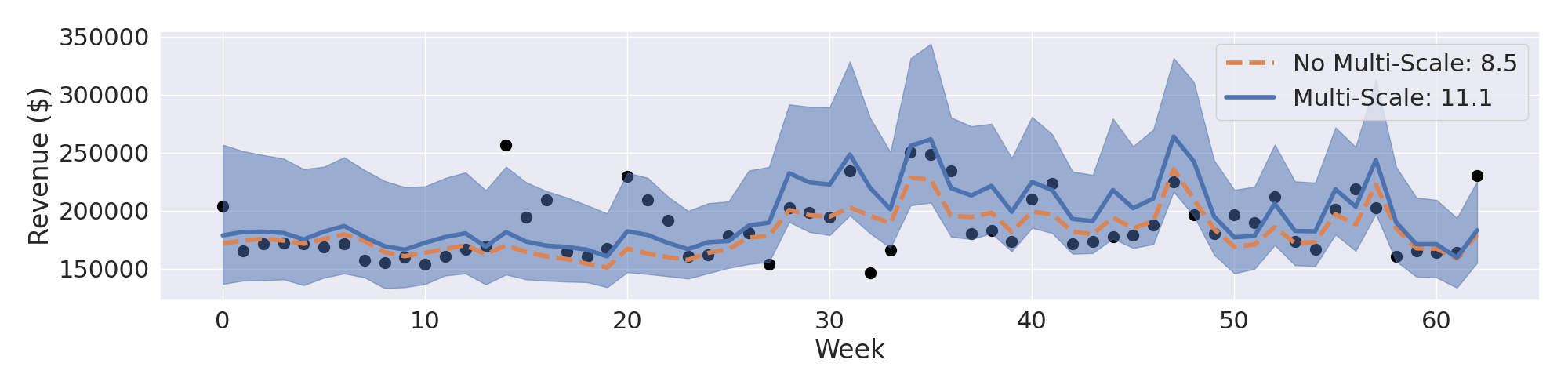}
   \caption{Forecasts - Local Store Group 3.}
\end{subfigure}
\begin{subfigure}[b]{0.95\textwidth}
   \includegraphics[width=1\linewidth]{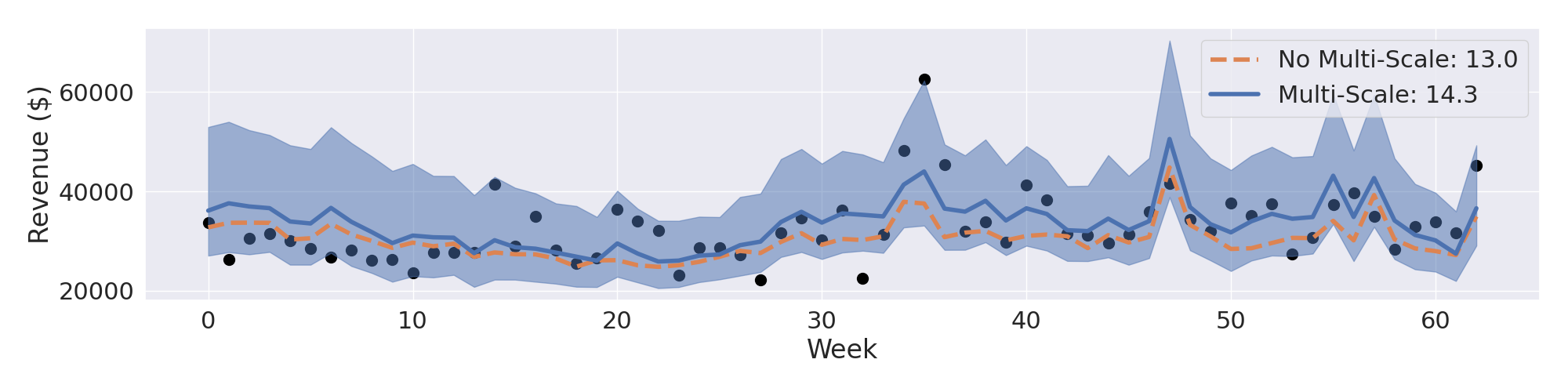}
   \caption{Forecasts - Local Store Group 6.}
\end{subfigure}
\begin{subfigure}[b]{0.95\textwidth}
   \includegraphics[width=1\linewidth]{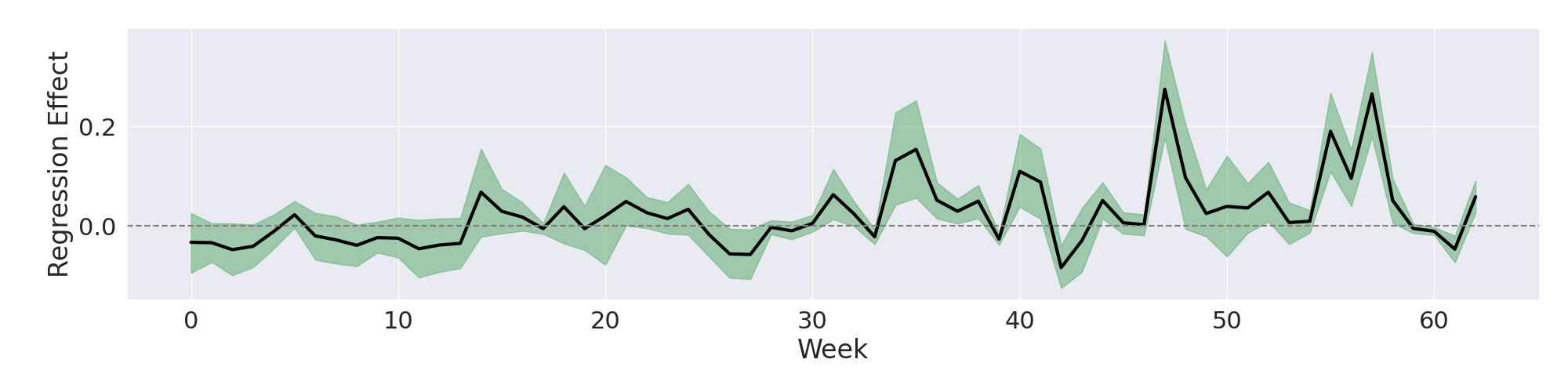}
   \caption{Regression Effect.}\label{fig:baked-reg}
\end{subfigure}
\caption{Summary 12-week ahead forecast and regression effect graphs for the Baked Sweet Goods Category for two LSGs. Details and format  as in \autoref{fig:sugars-main}.}
\label{fig:baked-main}
\end{figure}

\begin{figure}[!htp]
\centering
\begin{subfigure}[b]{0.95\textwidth}
   \includegraphics[width=1\linewidth]{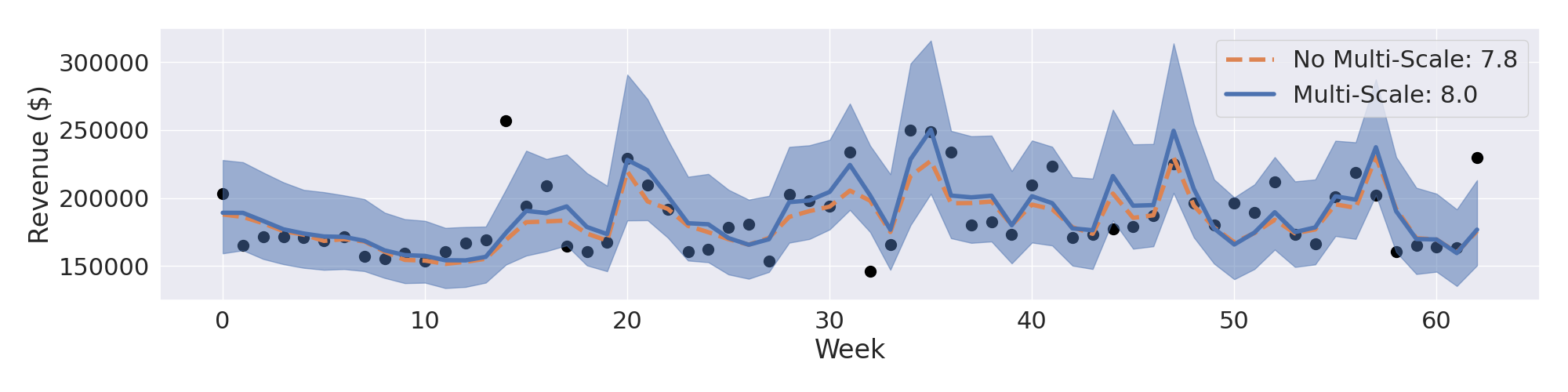}
   \caption{Forecasts - Local Store Group 3.}
\end{subfigure}
\begin{subfigure}[b]{0.95\textwidth}
   \includegraphics[width=1\linewidth]{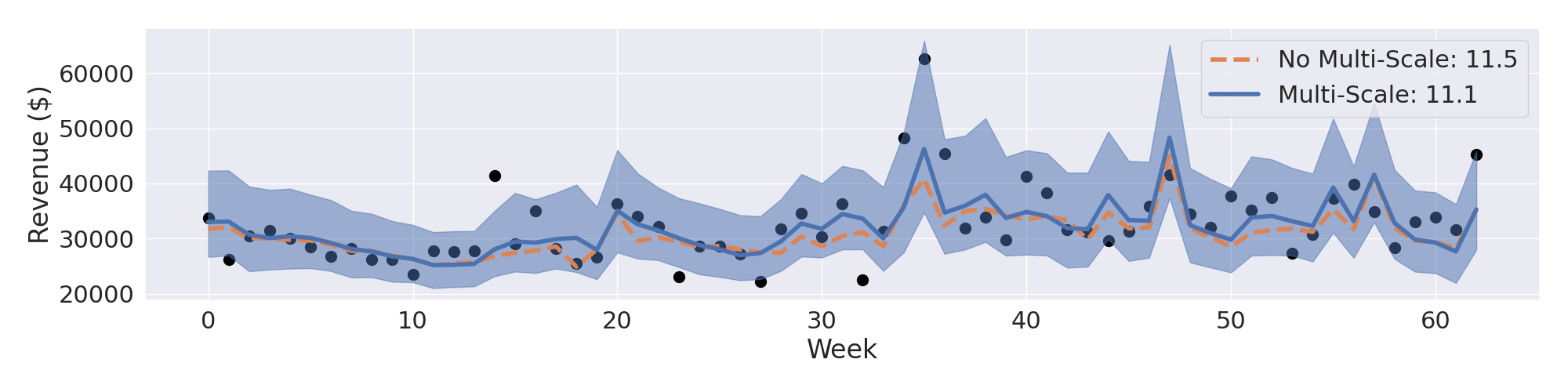}
   \caption{Forecasts - Local Store Group 6.}
\end{subfigure}
\caption{Summary 1 week ahead forecast graphs for the Baked Sweet Goods Category for two LSGs. Details and format as in \autoref{fig:sugars-k1}.}
\label{fig:baked-k1}
\end{figure}

Finally, Baked Sweet Goods is an example of a Category where multi-scale information does not improve revenue forecasts.  In general, from weeks 35-50, the multi-scale model tends to over-forecast, as reflected in both the forecasts themselves and the positive regression effect for this time period in \autoref{fig:baked-main}.  For weeks prior to week 35, the regression effect is approximately 0 and the no multi-scale and multi-scale models give very similar forecasts.  This Category could perhaps benefit from other types of multi-scale information that is more relevant, especially in early weeks when there is minimal discount regression effects.  One week ahead forecasts are given in \autoref{fig:baked-k1}.  

\FloatBarrier\newpage

\subsection{Extending the Revenue Models}\label{subsec:improve}

Additional information from the grocery chain offers potential to further improve  revenue forecasting in specific settings.  Here, we focus on the role of Category level pricing %, traffic 
and holiday effects.

\subsubsection{Pricing} 

There is additional information about pricing information via the Net Price variable; this is an average measure of the Net Price realized by customers (including discounts), averaged over customers within LSG and Category.  We find that jointly modeling and forecasting Net Price together with revenue can further improve revenue quite generally.  Updating the details  in Section~\ref{sec:methods}, the modifications are  as follows.  

Let $p_{t, c, z}$ be the Net Price for week $t$, Category $c$ and LSG $z$; we need to forecast $p_{t, c, z}$  as it incorporates realized discounts received by customers and so is uncertain in future weeks. We define a joint model by
coupling two univariate dynamic models: one for Net Price and one for 
revenue that extends the earlier DLM to also include Net Price as a predictor.  This decouple/recouple approach enables customization of each of the univariate model as well as sensitive modeling of dependence of revenue on Net Price. 
In summary, for each LSG-Category over weeks $t$ we: 
\begin{enumerate}[noitemsep,topsep=0pt]%
\item Model Net Price: $p_{t, c, z}\vert \bm{X}_{t, c, z}$.
\item Model revenue across LSGs (multi-scale): $Y_{t, c} \vert \bm{X}_{t, c}$.
\item Extract imputed values of the discount state vectors from model (2): $\bm{m}_{t, c}$ as before. 
\item Model revenue: $Y_{t, c, z}\vert \bm{X}_{t, c, z}, {p}_{t, c, z}, \bm{m}_{t, c}$
  now also conditional on imputed values of  ${p}_{t, c, z}.$
\end{enumerate}
At the final model stage, the imputed values of  ${p}_{t, c, z}$ can be any selected point forecasts;  the baseline choice is a \lq\lq plug-in" analysis that uses the forecast median of Net Price as from its univariate model.  This can be refined to run analyses repeatedly over a range of values or a Monte Carlo forecast sample of Net Price to understand if uncertainty under-quantification using the plug-in analysis is practically meaningful.  The revenue model also includes both the LSG-Category specific discount information and multi-scale discount information across LSGs, as before. The Net Price model uses the LSG-Category specific discount information and pricing information without discounts (the latter being which is a control variable for the grocery chain).

\begin{figure}[!ht]
\begin{center}
\includegraphics[width=\linewidth]{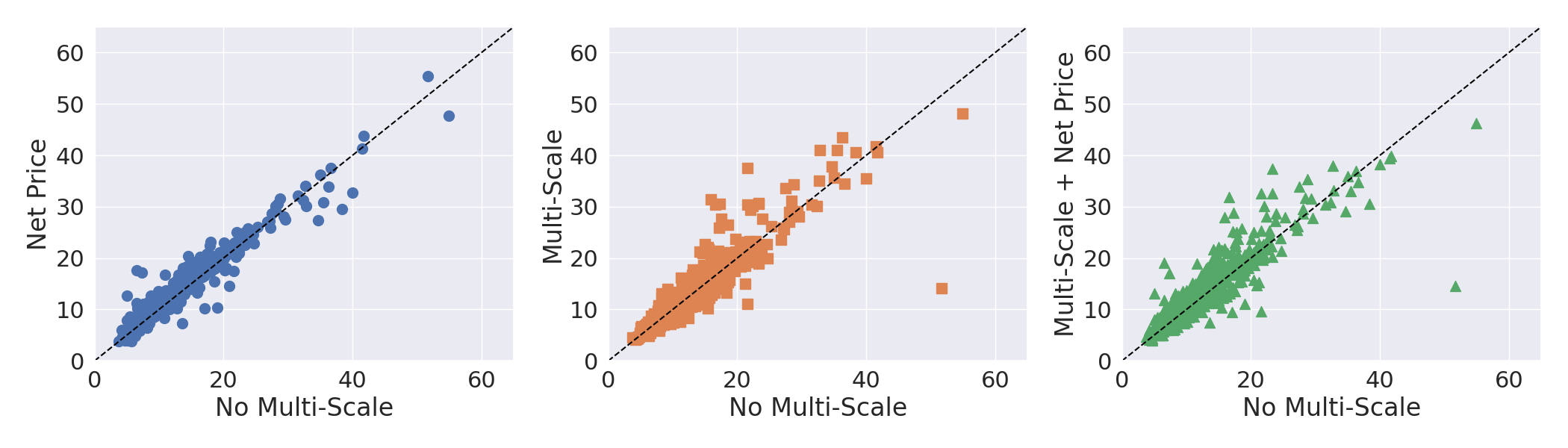}
 %\vspace*{2mm}
\caption{Empirical MAPE comparisons of revenue models versus baseline model. Each point represents the one-year average of 12-week ahead forecast MAPE for one LSG-Category pair.
}
\label{fig:scatter-net-compare}
\end{center}
\end{figure}

Selected aggregate results are highlighted in \autoref{fig:scatter-net-compare}.  With the set of univariate DLMs without multi-scale and Net Price extensions (\lq\lq No Multi-Scale") as baseline, this shows average revenue forecast MAPE values from (i) a revenue model with Net Price information only, (ii) the original multi-scale revenue model, and (iii) the more general revenue model with both multi-scale and Net Price information of this section.  Compared to the baseline, 
28\% of the LSG-Category pairs are improved with the Net Price model, 45\% for the multi-scale model, and 37\% for the multi-scale and Net Price model.  A number of specific LSG-Category pairs that particularly benefit from the inclusion of pricing information, while others do not. 

\begin{figure}[!ht]
\centering
\begin{subfigure}[b]{0.95\textwidth}
   \includegraphics[width=1\linewidth]{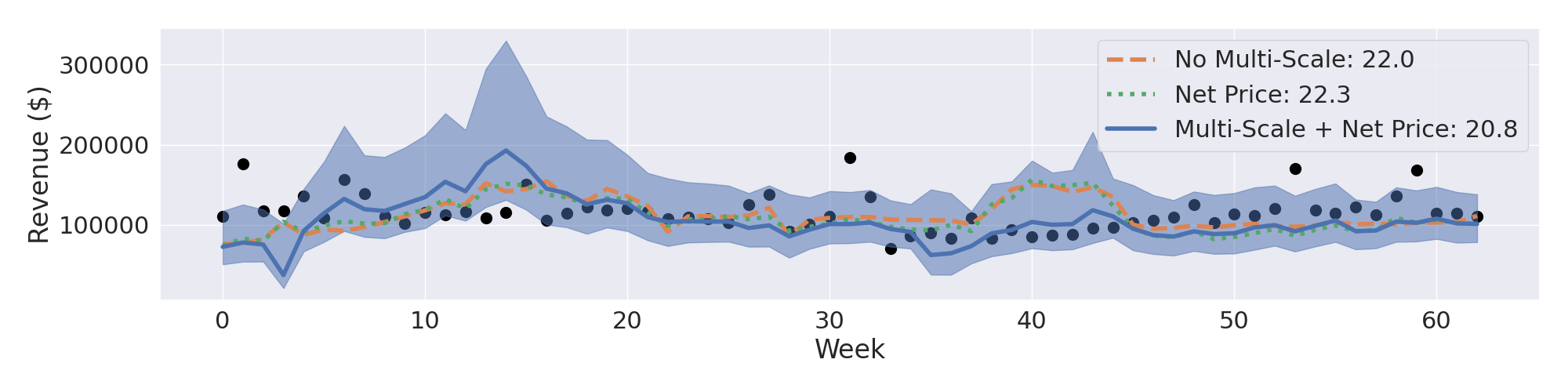}
   \caption{Forecasts - Local Store Group 3.}
\end{subfigure}
\begin{subfigure}[b]{0.95\textwidth}
   \includegraphics[width=1\linewidth]{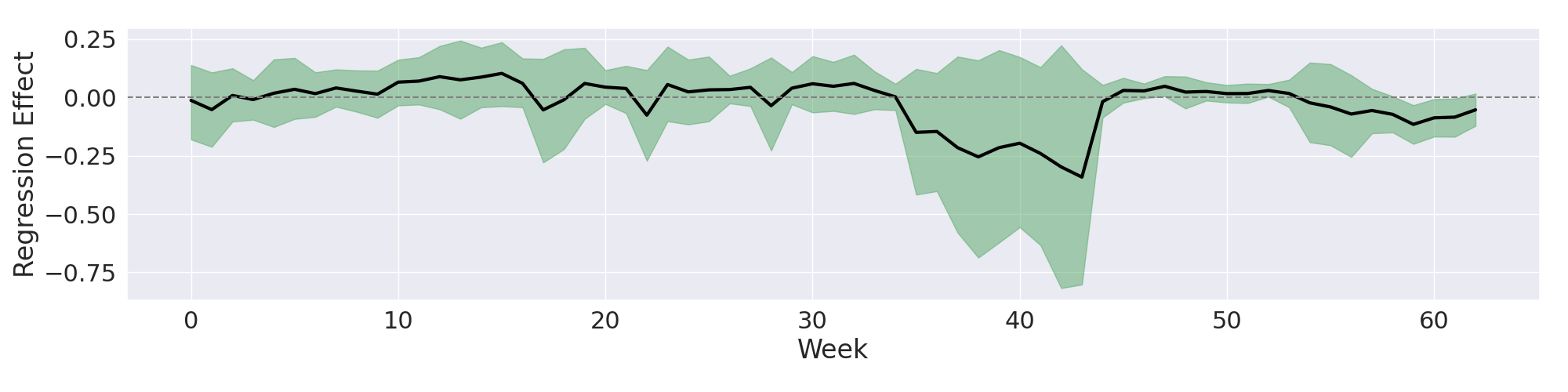}
   \caption{Regression Effect.}
\end{subfigure}
\caption{(a) Forecasts for the multi-scale, Net Price only, and multi-scale plus Net Price models (with MAPE values in the legend) for one LSG, and (b) estimated overall regression effect of predictors in the latter model. Format as in \autoref{fig:sugars-main}.}
\label{fig:beers-main}
\end{figure}

One Category where the combination of pricing and multi-scale discount information improves revenue forecasts is Craft/Micro Beers.  This Category is rarely discounted and when it is the discounts tend to be small.  There is also some retail price drift separate from discount information that can be helpful for this Category (see further comments in Supplementary Materials).  Forecast comparisons and regression effects are given in \autoref{fig:beers-main}.  The regression effect is generally insignificant over time.  We do see that, around weeks 35-40, the larger negative regression effect pulls down the forecasts in the multi-scale model, improving 12-week forecast accuracy over for this time period. While there is limited explanatory information in  LSG-specific or multi-scale discounts for this Category, they  nevertheless have practical value in revenue forecasting.

\begin{figure}[!ht]
\centering
\begin{subfigure}[b]{0.95\textwidth}
  \includegraphics[width=1\linewidth]{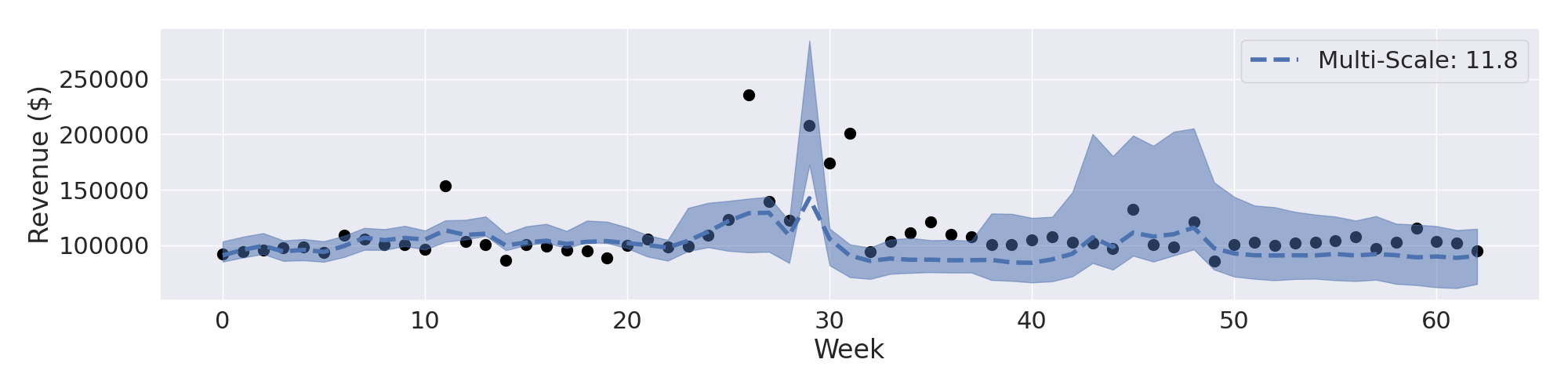}
  %\vskip1.3in{\color{red} Waiting for new fig from Anna}\vskip.5in
  \caption{Forecasts - Local Store Group 2.}
\end{subfigure}
\begin{subfigure}[b]{0.95\textwidth}
  \includegraphics[width=1\linewidth]{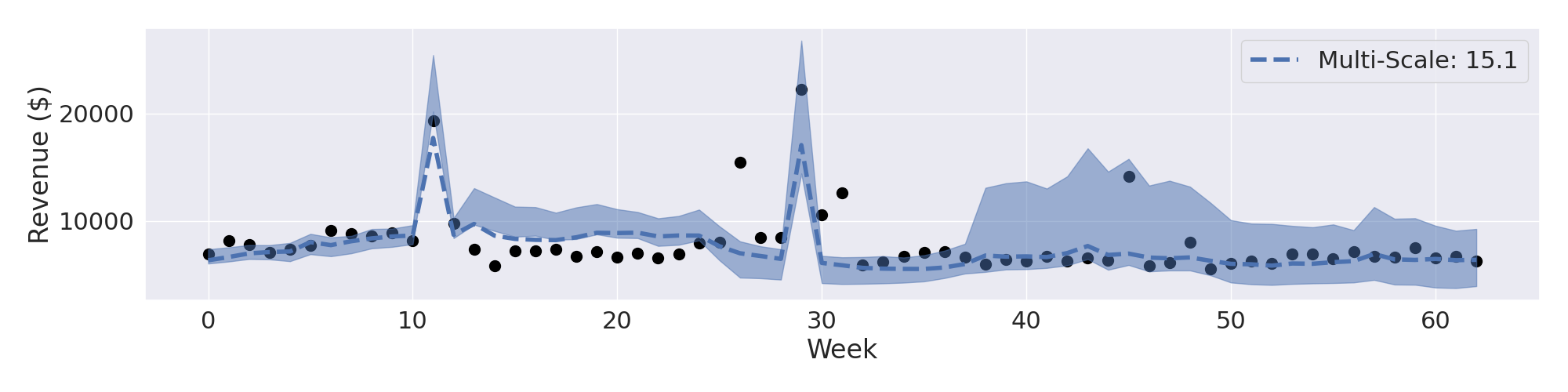}
  %\vskip1.3in{\color{red} Waiting for new fig from Anna}\vskip.5in
  \caption{Forecasts - Local Store Group 4.}
\end{subfigure}
\caption{12-week ahead revenue forecasts from the multi-scale model
for the Sugars \& Sweeteners Category in 2 LSGs. Format as in \autoref{fig:sugars-main}. }
\label{fig:rev-SugersNSweeteners}
\end{figure}

\subsubsection{Holiday Effects}

Some product Categories exhibit clear, important but sporadic holiday effects. The Sugars \& Sweeteners Category, for example, shows effects particularly around Christmas (\autoref{fig:rev-SugersNSweeteners}).  However, two years of data do not provide historical information sufficient to incorporate holiday week dummy variables, or holiday-specific transfer response model components over the week before, of and after the holiday period, such as is standard in Bayesian forecasting in commercial settings~\citep[][Sections 9.3 and 11.2]{West-Harrison}. Transfer response models designed specifically for local holiday effects have 
been utilized in related models in our setting, and coded for public access and incorporation into revenue models~\citep{PyBats}. 

The revenue models in further development for routine application are developed this way, but for our interest here we are mainly concerned about the impact of holiday events on forecast accuracy summaries. In terms of basic empirical accuracy impact, it is easy to re-evaluate MAPE (or other) metrics across all LSGs and Categories over the year of test data but simply dropping the (rare) holiday weeks from the summary.  This does not wholly re-evaluate accuracy, since the model analysis includes those weeks and so the sequential updating analysis is inevitably perturbed (negatively) by poor forecasts at holiday times that are not explicitly modeled as they might be, as noted above. But, simply masking out a few holiday weeks from the forecast error evaluation gives at least a lower bound on potential improvements.   

More formally, a fully Bayesian feed-forward intervention approach simply defines each holiday week as a known time when major departures from the routine model forecasts are expected, and treats the outcome data for those few weeks as missing observations.  This is effectively building in a \lq\lq holiday week" random intervention effect specific to each holiday, and with very high prior uncertainty. The result is that the state vectors in the baseline models will be protected from what may be large forecast errors in the forward filtering and updating analysis~~\citep[][Section  11.2.4]{BerryWest2018DCMM,West-Harrison}. 

Identifying the week of Thanksgiving, the week of Christmas and the week after Christmas (New Years') for the Sugars \& Sweeteners Category leads
%and re-run the multi-scale model with this weeks data ``missing''. %Results are given in \autoref{tab:holidays}.  
to strong aggregate improvements in terms of lower MAPE values; the net reduction in empirical MAPE values averaged over LSGs, Categories and across the 1 year evaluation period is about 7-8\%. This indicates that the three holiday periods have a substantial impact on forecast accuracy metrics. Some Categories are far more impacted than others, of course, and implementation of the models for routine use will customize developments for holidays as needed. For a subset of Categories, 
including specific holiday effects formally with more data is likely to be beneficial to  revenue forecasts.  This has been found to be the case internally by the grocery chain on separate data for which a longer period of time is available on some Categories and LSGs.

% \begin{table}[htp]
% \caption{Comparison of the MAPE median and standard deviation values across LSG-Category pairs for models with and without holidays included.  Both results are for multi-scale models.  The ``Holidays Included'' model is the same as the results given in DELETED TABLE REF % \autoref{tab:MAPE-models}.
% The  ``3 Holidays Dropped'' treats the weeks of Thanksgiving, Christmas and New Years' as missing to approximate holiday effects.  The performance improves with the holidays approximately accounted for.}
%  %\vspace*{5mm}
% \begin{center}
% \begin{tabular}{c|c}
% Setting & MAPE \\\hline
% Holidays Included & $10.61\pm 6.05$ \\
% 3 Holidays Dropped & $9.81\pm 4.81$ \\
% \end{tabular}
% \end{center}
% \label{tab:holidays}
% \end{table}%

\subsection{Exploration of Cross-Category Dependence}\label{subsec:cross-cat}

There are business interests in identifying whether discounts for one Category affect sales and hence revenue in other Categories. Identifying such relationships has potential to yield forecast accuracy improvements by including  relevant cross-Category discount predictors in revenue models.  Then, if higher discounts for Category A are associated with higher sales for Category B,  the  products within the Categories are potential complements and cross-Category promotion strategies may be of interest to management. A store or LSG could offer discounts in Category A to induce customers to also purchase products in Category B at lesser discounts. On the other hand, if higher discounts for Category A are associated with lower sales for Category B, then products within the two Categories are possible substitutes of each other, and discounts potentially \lq\lq cannibalize" cross-Category sales. If some products in Category A are heavily discounted and sales within Category B decrease, then consumption has merely shifted and apparent sales lift in Category A is masking potentially store-level, or LSG-level, drops in revenue. 

We identify potential pairs for cross-Category analysis by examining relationships between standardized forecast errors from the log revenue models. This is exemplified here using the 12-week multi-scale revenue  models incorporating  multi-scale discount, Net Price, and the primary discount variables as predictors. Post-forecasting exploration of 12-week ahead forecast errors is key as these realized errors are implicitly already free (modulo the assumed adequacy of the models) of the effects of Category-specific discounts and other effects that may generate spurious indications of cross-Category relationships.  The later include, for example, any patterns of local trend and/or seasonality that may be common to Category revenue and discount decisions; e.g. sales of hot chocolate increase in the winter while discounts on ice cream decrease.  Further, we use realized errors standardized under their step ahead forecast distributions; this appropriately accounts for series-specific residual volatility over time prior to evaluating cross-Category correlations.

It is also important to examine  consistency of any potential cross-Category relationships across Local Store Groups. Each LSG has, in theory, the ability to independently select discount strategies in any Category for stores in the LSG. If an observed cross-Category relationship is consistent across LSGs, then that Category pair is of more interest for further exploration.   

% Some summaries of exploratory analysis using all $40\times 40$ Category combinations are highlighted. 
% Figure~\ref{fig:correlation_mats_revenue} presents a heat-map of cross-Category correlations of revenue with TPR\% discount. Each $i,j$ entry is the empirical correlation between revenue in Category $i$ and TPR\% in Category $j$ evaluated over the 52-week forecasting test period and averaged across the 9 LSGs.
% This naive display shows many apparently interesting correlations that are spurious; they disappear when evaluation uses forecast errors instead of revenue; see Figures~\ref{fig:correlation_mats_errorsk1} and~\ref{fig:correlation_mats_errorsk12}.   Analysis using 12-week forecast errors shows evidence of some stronger correlations than using 1-week forecast errors. This is important since the 12-week horizon is 
% most relevant for business decisions; at that horizon, forecasts are generally less accurate than the 1-week forecasts, so there is more room for improvement by incorporating cross-Category promotion strategies in the 12-week forecasting models.  

\begin{figure}[htbp!]
\centering
\includegraphics[width=0.95\textwidth]{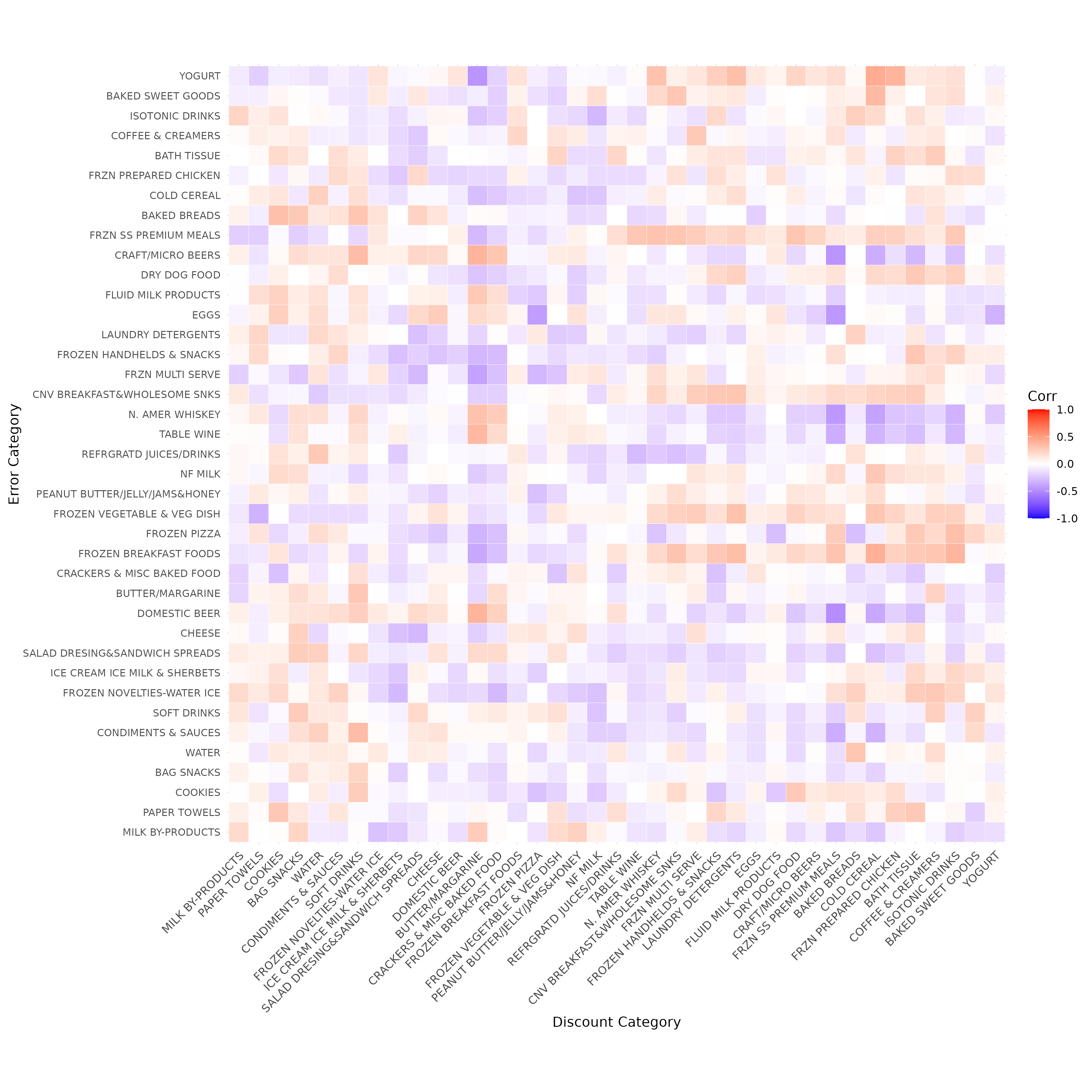}
\caption[Correlations Between 12-week Errors  and cross-Category Discounts.]{Heat-map of correlations between realized 12-week ahead Revenue forecast errors and TPR\% discounts. The $i,j$ entry shows the empirical correlation between the forecast errors in Category $i$ and TPR\% (the percentage of products on discount) in Category $j$. These are evaluated over the 52-week forecasting test period and then averaged across all 9 LSGs.
}
\label{fig:correlation_mats_errorsk12}
\end{figure}

Some summaries of exploratory analysis using the top $40\times 40$ Category combinations are highlighted. 
Figure~\ref{fig:correlation_mats_errorsk12} presents a heat-map of cross-Category correlations of forecast errors from the 12-week ahead multi-scale revenue models with TPR\% discount. For each pair of Categories $i,j$ the correlation is that between realized forecast errors in Category $i$ and TPR\% in Category $j$ 
evaluated over the 52-week forecasting test period and averaged across the 9 LSGs.
While many pairwise correlations are apparently negligible, interest lies in exploring specific example pairs where the correlation seems highest.  First note, however, that the
corresponding correlation heat-map based on raw revenue data rather than on the model-based forecast errors shows substantial numbers of much higher correlations (Supplementary Material, Figure~\ref{fig:correlation_mats_revenue}).  The naive analysis using raw revenue generates many  
apparently interesting  but spurious suggestions of cross-Category relationships that disappear when evaluation uses forecast errors instead of revenue.  A more important comparison is with the corresponding heat-map of correlations using using 1-week rather than 12-week ahead forecast errors
(Supplementary Material, Figure~\ref{fig:correlation_mats_errorsk1}). Analysis at the longer forecast horizon  shows evidence of some stronger correlations than using 1-week forecast errors. This is important since the 12-week horizon is most relevant for business decisions; at that horizon, forecasts are generally less accurate than the 1-week forecasts, so there is more room for improvement by incorporating cross-Category promotion strategies in the 12-week forecasting models.  

\begin{figure}[!ht]
\begin{center}
\includegraphics[width=1\linewidth]{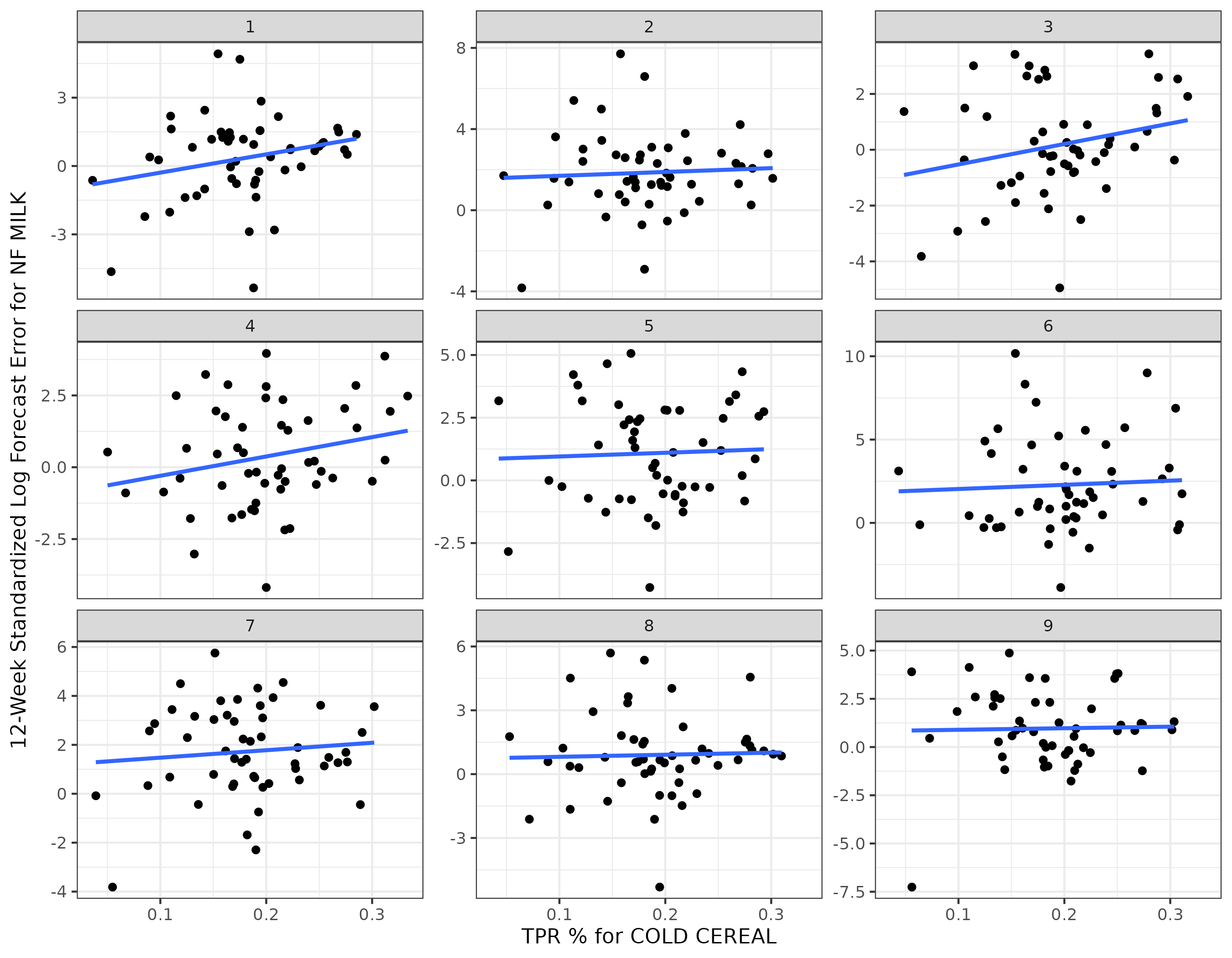}
 \vspace*{2mm}
\caption[Relationship between forecast errors and TPR percent.]{Scatter plots of 12-week ahead forecast errors for NF Milk revenue against the TPR\% discount measure for Cold Cereal products across the 52 week test period, for each of the 9 LSGs.   }
\label{fig:cereal_milk_scatter}
\end{center}
\end{figure}

We highlight one particular pair of Categories: Cold Cereal and NF (organic) Milk. Discounts for Cold Cereal have the largest correlation with NF Milk forecast errors across the examined Categories in two of the nine LSGs, always positive, and fourth largest when averaged across LSGs. Figure \ref{fig:cereal_milk_scatter} shows  12-week forecast errors for NF Milk against Cold Cereal TPR\% discount for each of the LSGs.  Slightly positive-- albeit rather weak and noisy-- relationships are consistent with the view that the models tend to under-predict NF Milk revenues when Cold Cereal experiences higher discounts.  The concordance across several LSGs is important in supporting the view that this is a systematic, potentially casual relationship. From a forecasting viewpoint,  the potential for such a cross-Category association to be useful is explored by  re-estimating the revenue for NF Milk but now extending the model to also include the Cold Cereal TPR\% discount as a predictor.  That analysis was performed and confirmed  point forecast improvements; the 12-week ahead MAPE metric averaged over the 52 week test period is reduced for 6 of the 9 LSGs and remains essentially  unchanged for the other 3. Again, we repeat the point that even very small improvements in this measure of forecast accuracy at the LSG level can be of real practical importance in informing planning, promotion and logistics with meaningful business revenue impact.  
%Figure \ref{fig:cereal_milk_mape} shows MAPE loss is lower or roughly the same when including Cold Cereal discounts as a predictor for NF Milk sales. 

% \begin{figure}[!ht]
% \begin{center}
% \includegraphics[width=\linewidth]{\figdir/cross-category/cereal_discount_impact.png}
%  \vspace*{2mm}
% \caption[Reductions in MAPE after including cross-Category discounts.]{Each point shows the MAPE loss for NF Milk forecasts in an individual LSG averaged across the final year of data, either including Cold Cereal discounts as a predictor (y-axis) or excluding them (x-axis). Points below the line indicate MAPE is reduced by including the cross-Category discounts.}
% \label{fig:cereal_milk_mape}
% \end{center}
% \end{figure}

\FloatBarrier\newpage

%%%%%%%%%%%%%%%%%%%%%%%%%%%%%%%
\section{Summary Comments}\label{sec:conc}

Our case study of revenue modeling at the LSG-Category level for a large grocery chain has extended Bayesian multi-scale dynamic modeling to enable integration of series specific as well as cross-Category discount information in forecasting several hundred multivariate revenue time series. A few summaries here of the much broader analysis exhibit key practical aspects of these joint models for pricing and revenue, and examine features of cross-Category dependencies.  Substantial heterogeneity across both LSGs and Categories offers opportunities for multi-scale, aggregate information sharing to improve LSG-Category specific forecasts.  The multi-scale signal in this setting is the aggregate state vector information related to discounts for each Category across LSGs, and we find that this can improve multi-step ahead (12-week ahead) forecasts  for about half of the main Categories of interest to the company.  The baseline dynamic models should be maintained for the other Categories and they already define forecasting advances in relying on LSG-Category predictors generated from pricing and discount information, as well as benefiting from the inherent adaptability over time of Bayesian DLMs. For the LSG-Category cases that do benefit from the multi-scale extension, forecast improvements are practically relevant and some quite large in terms of revenue implications.   

There are several avenues for future development in applications and methodology.  The company is involved in developing broader evaluation on more extensive data sets including explicit integration of holiday effects in the DLMs.   Additional exploration of other types of multi-scale information, for example across groups of similar Categories, is one direction that raises potential for further improvements in forecast accuracy.  In particular, alcohol (and other) Categories that are rarely discounted are likely to benefit from information sharing across multiple contextually-related Categories, in addition to across LSGs.  Additionally, measures of traffic, such as weekly transactions within an LSG containing items of a given category, might be jointly modeled with category revenue to improve forecast accuracy.  Exploring cross-Category dependence is possible with this modeling approach and is of interest for the application, specifically to understand how discounts in one Category impact revenue in another Category.  This ties into more formal ``What-if?'' decision analyses-- also known as ``scenario forecasting''-- to explore, for example, how changes in pricing or promotions for specific Categories leads to changes in revenue in the same or other Categories.  The potential for extending this line of thinking to a causal basis, involving real-time experimentation, is clearly an open and interesting area, though as yet not a direction addressed in public-domain R\&D linked to this specific study. 

We note that the model analyses presented and summarized can be developed by interested readers and potential users based on prototype code available in PyBats~\citep{PyBats}.

%%%%%%%%%%%%%%%%%%%%%%%%%%%%%%%%%%%%%%%%%%%%%%%%%

%\section*{Acknowledgements}

%The research reported here was partly supported by $84.51^\circ$. Our research has benefited from discussions with  $84.51^\circ$ Research Scientist Christoph Hellmayr.  Any opinions, findings and conclusions or recommendations expressed in this paper do not necessarily reflect the views of $84.51^\circ$. 

%%%%%%%%%%%%%%%%%%%%%%%%%%%%%%%%%%%%%%%%%%%%%%%%
%% References

%\bibliographystyle{myplainnat}
\bibliographystyle{chicago} % Style BST file
\bibliography{BayesianRevenueForecasting_References}

\begin{thebibliography}{}

\bibitem[\protect\citeauthoryear{Berry}{Berry}{2019}]{Berry:2019}
Berry, L.~R. (2019).
\newblock {\em Bayesian Dynamic Modeling and Forecasting of Count Time Series}.
\newblock Ph.\ D. thesis, Department of Statistical Science, Duke University.

\bibitem[\protect\citeauthoryear{Berry, Helman, and West}{Berry
  et~al.}{2020}]{BerryWest2018DBCM}
Berry, L.~R., P.~Helman, and M.~West (2020).
\newblock Probabilistic forecasting of heterogeneous consumer transaction-sales
  time series.
\newblock {\em International Journal of Forecasting\/}~{\em 36}, 552--569.

\bibitem[\protect\citeauthoryear{Berry and West}{Berry and
  West}{2020}]{BerryWest2018DCMM}
Berry, L.~R. and M.~West (2020).
\newblock Bayesian forecasting of many count-valued time series.
\newblock {\em Journal of Business and Economic Statistics\/}~{\em 38},
  872--887.

\bibitem[\protect\citeauthoryear{Chu and Zhang}{Chu and Zhang}{2003}]{Chu:2003}
Chu, C.-W. and G.~P. Zhang (2003).
\newblock A comparative study of linear and nonlinear models for aggregate
  retail sales forecasting.
\newblock {\em International Journal of Production Economics\/}~{\em 86},
  217--231.

\bibitem[\protect\citeauthoryear{Lavine and Cron}{Lavine and
  Cron}{2020}]{PyBats}
Lavine, I. and A.~Cron (2020).
\newblock {PyBats}: A {P}ython package for \underbar {B}ayesian \underbar
  {A}nalysis of \underbar {T}ime \underbar {S}eries and {B}ayesian forecasting.
\newblock \url{https://pypi.org/project/pybats/}.

\bibitem[\protect\citeauthoryear{Lei and Cailan}{Lei and
  Cailan}{2021}]{Lei:2021}
Lei, H. and H.~Cailan (2021).
\newblock Comparison of multiple machine learning models based on enterprise
  revenue forecasting.
\newblock In {\em 2021 Asia-Pacific Conference on Communications Technology and
  Computer Science (ACCTCS)}, pp.\  354--359.

\bibitem[\protect\citeauthoryear{Mishev, Gjorgjevikj, Vodenska, Chitkushev,
  Souma, and Trajanov}{Mishev et~al.}{2019}]{Mishev:2019}
Mishev, K., A.~Gjorgjevikj, I.~Vodenska, L.~Chitkushev, W.~Souma, and
  D.~Trajanov (2019).
\newblock Forecasting corporate revenue by using deep-learning methodologies.
\newblock In {\em 2019 International Conference on Control, Artificial
  Intelligence, Robotics Optimization (ICCAIRO)}, pp.\  115--120.

\bibitem[\protect\citeauthoryear{Prado, Ferreira, and West}{Prado
  et~al.}{2021}]{PradoFerreiraWest2021}
Prado, R., M.~A.~R. Ferreira, and M.~West (2021).
\newblock {\em Time Series: Modeling, Computation \& Inference\/} (2nd ed.).
\newblock Chapman \& Hall/CRC Press.

\bibitem[\protect\citeauthoryear{Pundir, Ganapathy, Maheshwari, and
  Kumar}{Pundir et~al.}{2020}]{Pundir:2020}
Pundir, A.~K., L.~Ganapathy, P.~Maheshwari, and M.~N. Kumar (2020).
\newblock Machine learning for revenue forecasting: A case study in retail
  business.
\newblock In {\em 2020 11th IEEE Annual Information Technology, Electronics and
  Mobile Communication Conference (IEMCON)}, pp.\  201--207.

\bibitem[\protect\citeauthoryear{Salinas, Bohlke-Schneider, Callot, Medico, and
  Gasthaus}{Salinas et~al.}{2019}]{NIPS2019_8907}
Salinas, D., M.~Bohlke-Schneider, L.~Callot, R.~Medico, and J.~Gasthaus (2019).
\newblock High-dimensional multivariate forecasting with low-rank {G}aussian
  copula processes.
\newblock In {\em Advances in Neural Information Processing Systems},
  Volume~32, pp.\  6827--6837. Curran Associates, Inc.

\bibitem[\protect\citeauthoryear{Sen, Yu, and Dhillon}{Sen
  et~al.}{2019}]{sen2019think}
Sen, R., H.-F. Yu, and I.~S. Dhillon (2019).
\newblock Think globally, act locally: A deep neural network approach to
  high-dimensional time series forecasting.
\newblock In {\em Advances in Neural Information Processing Systems},
  Volume~32. Curran Associates, Inc.

\bibitem[\protect\citeauthoryear{Weatherford}{Weatherford}{2016}]{Weatherford:2016}
Weatherford, L. (2016).
\newblock The history of forecasting models in revenue management.
\newblock {\em Journal of Revenue and Pricing Management\/}~{\em 15}, 212--221.

\bibitem[\protect\citeauthoryear{West}{West}{2020}]{West2020Akaike}
West, M. (2020).
\newblock Bayesian forecasting of multivariate time series: {S}calability,
  structure uncertainty and decisions (with discussion).
\newblock {\em Annals of the Institute of Statistical Mathematics\/}~{\em 72},
  1--44.

\bibitem[\protect\citeauthoryear{West and Harrison}{West and
  Harrison}{1997}]{West-Harrison}
West, M. and P.~J. Harrison (1997).
\newblock {\em Bayesian Forecasting and Dynamic Models\/} (2nd ed.).
\newblock Springer-Verlag, New York, Inc.

\bibitem[\protect\citeauthoryear{Yanchenko, Deng, Li, Cron, and West}{Yanchenko
  et~al.}{2021}]{yanchenko2021hierarchical}
Yanchenko, A.~K., D.~Deng, J.~Li, A.~J. Cron, and M.~West (2021).
\newblock Hierarchical dynamic modeling for individualized {B}ayesian
  forecasting (submitted).
\newblock arXiv: 2101.03408.

\end{thebibliography}
 
\newpage
%\appendix
\setcounter{page}{1}

\begin{center}
    {\bf\Large Multivariate Dynamic Modeling for \\ Bayesian Forecasting of Business Revenue \\ 
     \phantom{.}\\ -- Supplementary Material --}

\bigskip\bigskip
Anna K. Yanchenko\footnote{Department of Statistical Science,Duke University, Durham NC 27708-0251. U.S.A.}, Graham Tierney$^1$, Joseph Lawson$^1$, Christoph Hellmayr\footnote{$84.51^\circ$, 100 West 5th Street, Cincinnati, OH 45202.  U.S.A.}, 
\\ Andrew Cron$^2$ \& Mike West$^1$

\bigskip\today\bigskip

\end{center}

\noindent This Supplement highlights additional examples and summary details of revenue forecast results as well as cross-Category dependence analysis.

\subsection*{Additional Summaries of Aggregate Level Forecasting Results}

Full dynamic model forecasting analyses across all $9\times 100$ LSG-Category pairs define MAPE-based assessment of point forecasting accuracy for comparison of chosen models. These analyses compare: (i) the baseline univariate models using local trend, seasonality and LSG-Category specific price discount predictors; (ii) the extended univariate models that also include the Net Price predictor and that involve/require also modeling and forecasting Net Price; (iii) the baseline univariate models coupled via the multi-scale extension that involves modeling and forecasting of the aggregate level, cross-LSG discount effects per Category; and (iv) the combination of model forms (ii) and (iii) into a multi-scale plus Net Price model form.  

Relative to the baseline univariate models, the empirical results using MAPE averaged over the 52 week test period and across LSGs show the following. On either 1-week ahead or 12-week ahead forecasting assessments: 
\begin{itemize}[noitemsep,topsep=0pt]%
\item adding Net Price alone improves forecast accuracy for 31\% (1-week) and 28\% (12-week) of the LSG-Category pairs; 
\item extending to include multi-scale discount predictors improves forecast accuracy for 62\% (1-week) and 45\% (12-week) of the LSG-Category pairs; 
\item adding Net Price and the multi-scale extension improves forecast accuracy for 52\% (1-week) and 37\% (12-week) of the LSG-Category pairs. 
\end{itemize} 
More LSG-Category pairs benefit from multi-scale information at the 1-week horizon.

\newpage
\subsection*{Additional LSG-Category Examples}

\begin{figure}[htbp!]
\centering
\begin{subfigure}[b]{0.95\textwidth}
   \includegraphics[width=1\linewidth]{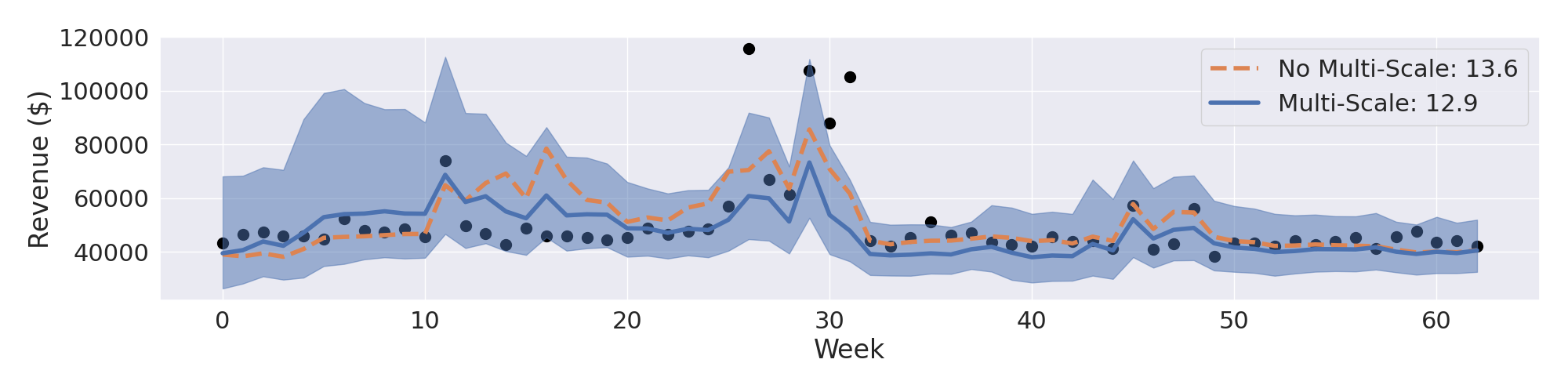}
   \caption{Forecasts - Local Store Group 1.}
\end{subfigure}
\begin{subfigure}[b]{0.95\textwidth}
   \includegraphics[width=1\linewidth]{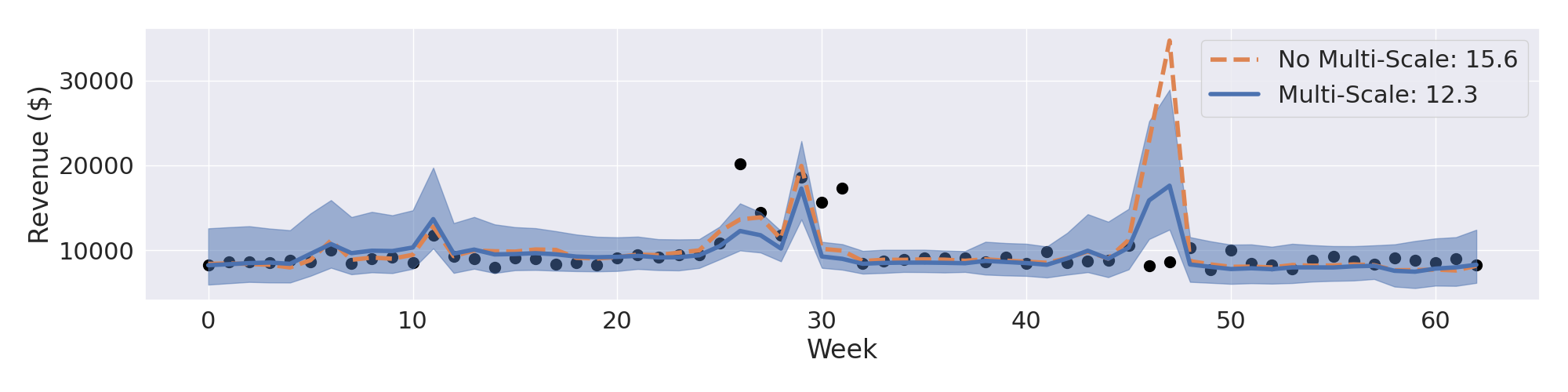}
   \caption{Forecasts - Local Store Group 9.}
\end{subfigure}
\begin{subfigure}[b]{0.95\textwidth}
   \includegraphics[width=1\linewidth]{{"\figdir/revenue/DIV=014-LSG=None-CAT-15-SUGARS_SWEETENERS-mt-MS-regression-k12"}.png}
   \caption{Regression Effect.}
\end{subfigure}
\caption{12-week ahead forecasts and discount regression effects for the multi-scale and no multi-scale model for the Sugar \& Sweeteners Category for two LSGs. Format as in \autoref{fig:sugars-main}.}
\label{fig:sugars-supp}
\end{figure}

\begin{figure}[htbp!]
\centering
\begin{subfigure}[b]{0.95\textwidth}
   \includegraphics[width=1\linewidth]{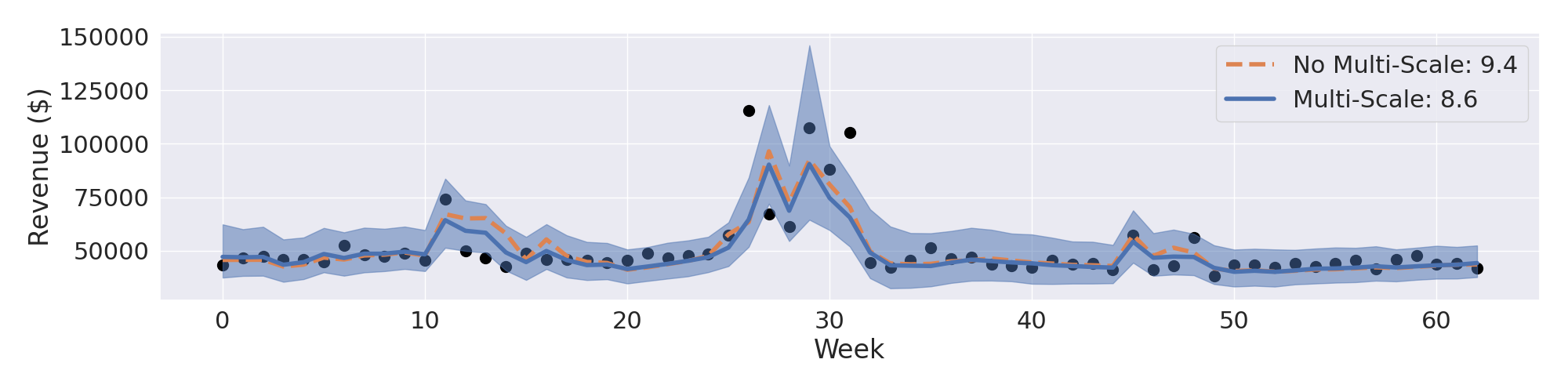}
   \caption{Forecasts - Local Store Group 1.}
\end{subfigure}
\begin{subfigure}[b]{0.95\textwidth}
   \includegraphics[width=1\linewidth]{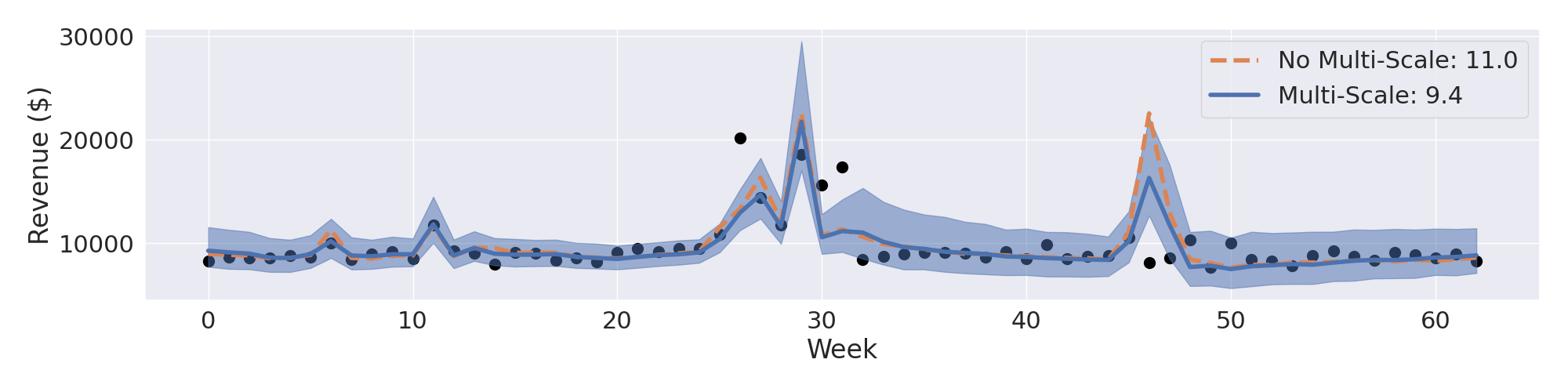}
   \caption{Forecasts - Local Store Group 9.}
\end{subfigure}
\caption{1-week ahead forecasts and discount regression effects for the multi-scale and no multi-scale model for the Sugar \& Sweeteners Category for two LSGs. Format as in \autoref{fig:sugars-main}.}
\label{fig:sugars-supp-k1}
\end{figure}

\begin{figure}[htbp!]
\centering
\begin{subfigure}[b]{0.95\textwidth}
   \includegraphics[width=1\linewidth]{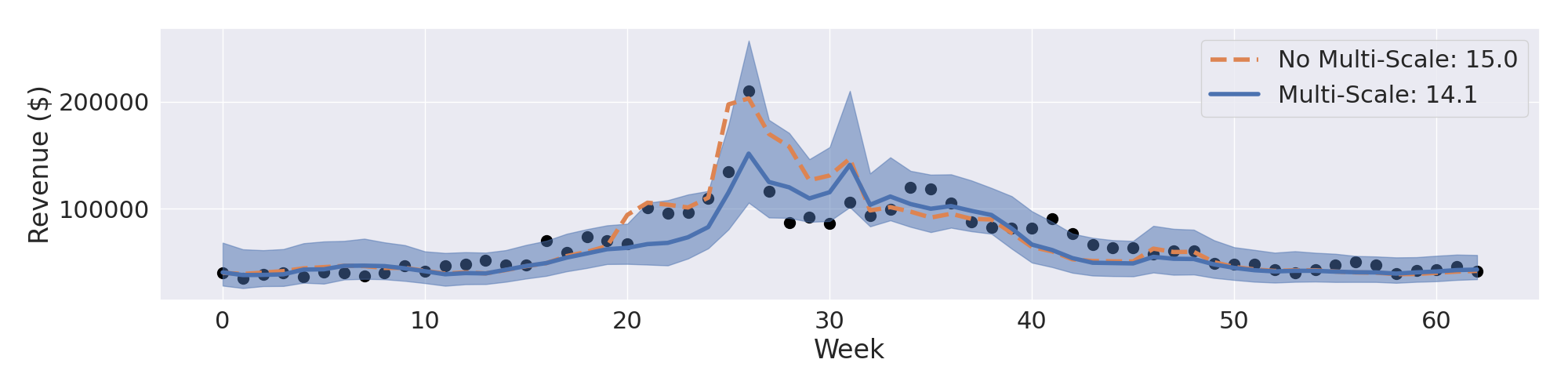}
   \caption{Forecasts - Local Store Group 2.}
\end{subfigure}
\begin{subfigure}[b]{0.95\textwidth}
   \includegraphics[width=1\linewidth]{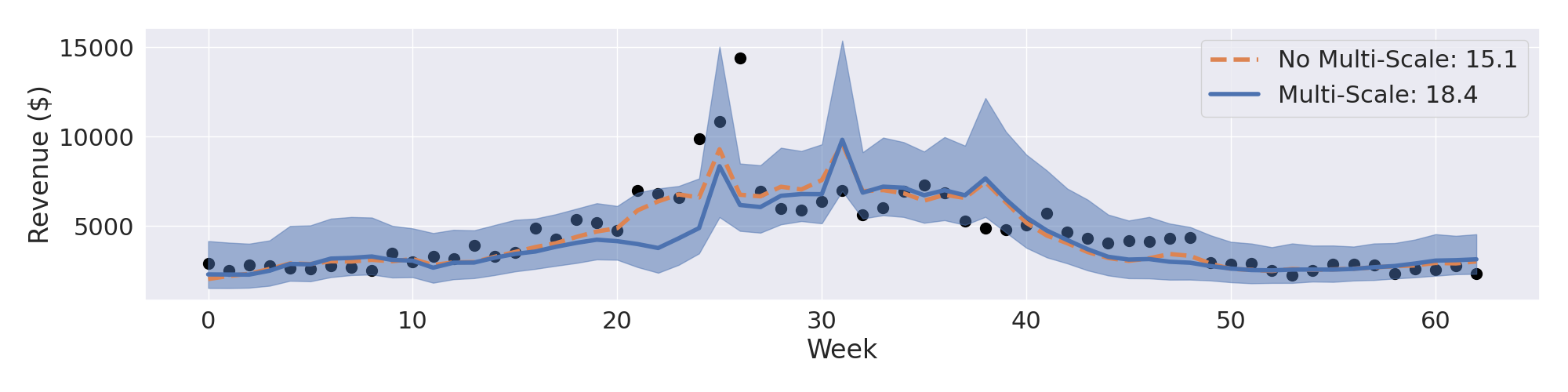}
   \caption{Forecasts - Local Store Group 4.}
\end{subfigure}
\begin{subfigure}[b]{0.95\textwidth}
   \includegraphics[width=1\linewidth]{{"\figdir/revenue/DIV=014-LSG=None-CAT-154-BROTH_BOUILLON-mt-MS-regression-k12"}.png}
   \caption{Regression Effect.}
\end{subfigure}
\caption{12-week ahead forecasts and discount regression effects for the multi-scale and no multi-scale model for the Broth/Dry Soup Category for two LSGs. Format as in \autoref{fig:sugars-main}.}
\label{fig:broth-supp}
\end{figure}

\begin{figure}[htbp!]
\centering
\begin{subfigure}[b]{0.95\textwidth}
   \includegraphics[width=1\linewidth]{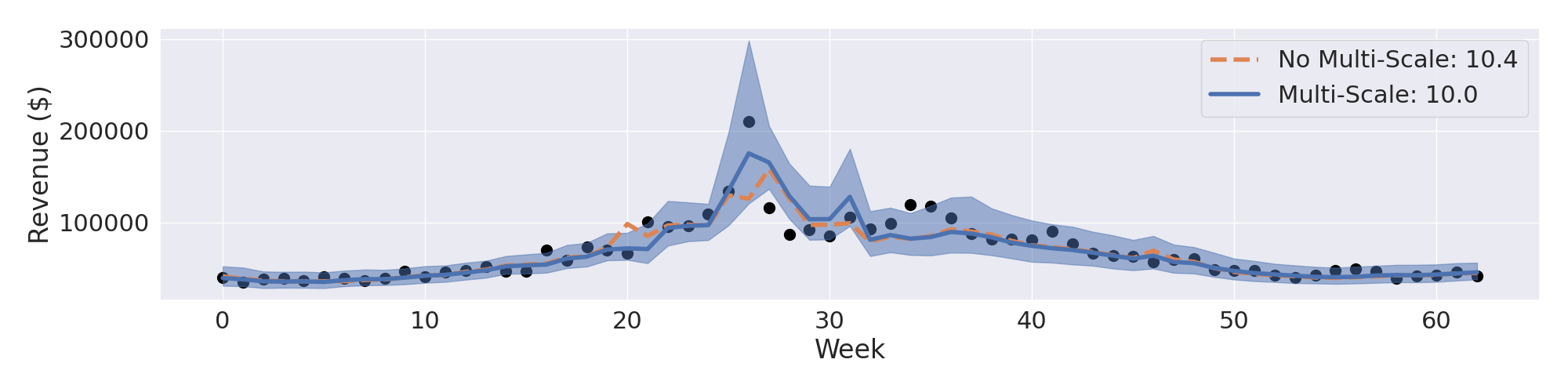}
   \caption{Forecasts - Local Store Group 2.}
\end{subfigure}
\begin{subfigure}[b]{0.95\textwidth}
   \includegraphics[width=1\linewidth]{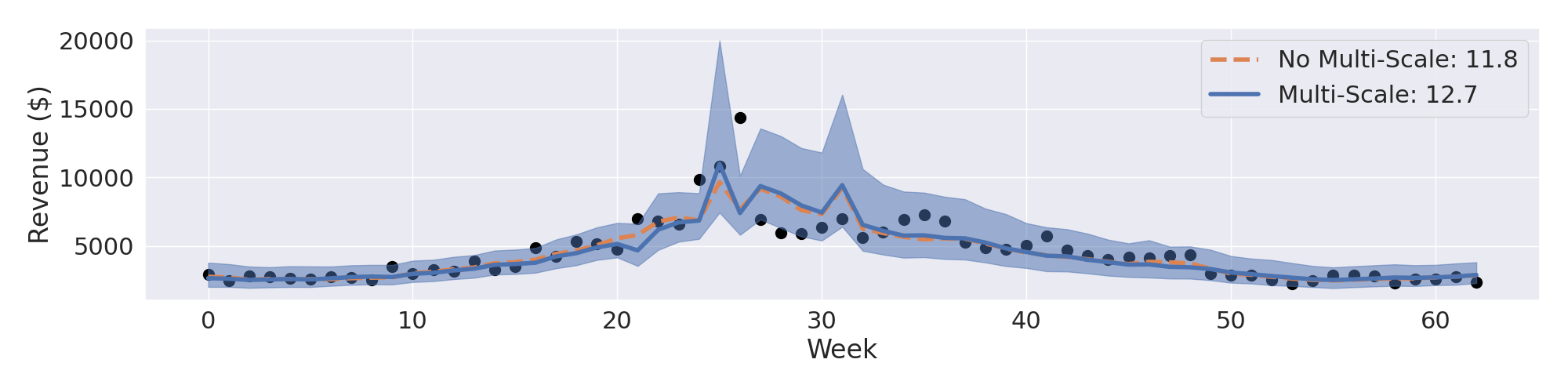}
   \caption{Forecasts - Local Store Group 6.}
\end{subfigure}
\caption{1-week ahead forecasts and discount regression effects for the multi-scale and no multi-scale model for the Broth/Dry Soup Category for two LSGs. Format as in \autoref{fig:sugars-main}.}
\label{fig:broth-supp-k1}
\end{figure}

\begin{figure}[htbp!]
\centering
\begin{subfigure}[b]{0.95\textwidth}
   \includegraphics[width=1\linewidth]{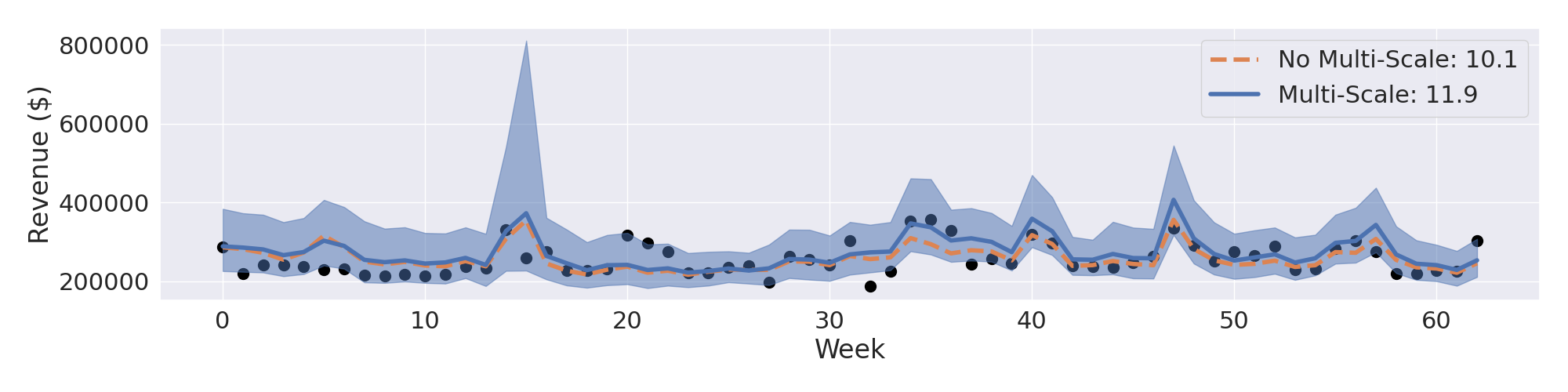}
   \caption{Forecasts - Local Store Group 7.}
\end{subfigure}
\begin{subfigure}[b]{0.95\textwidth}
   \includegraphics[width=1\linewidth]{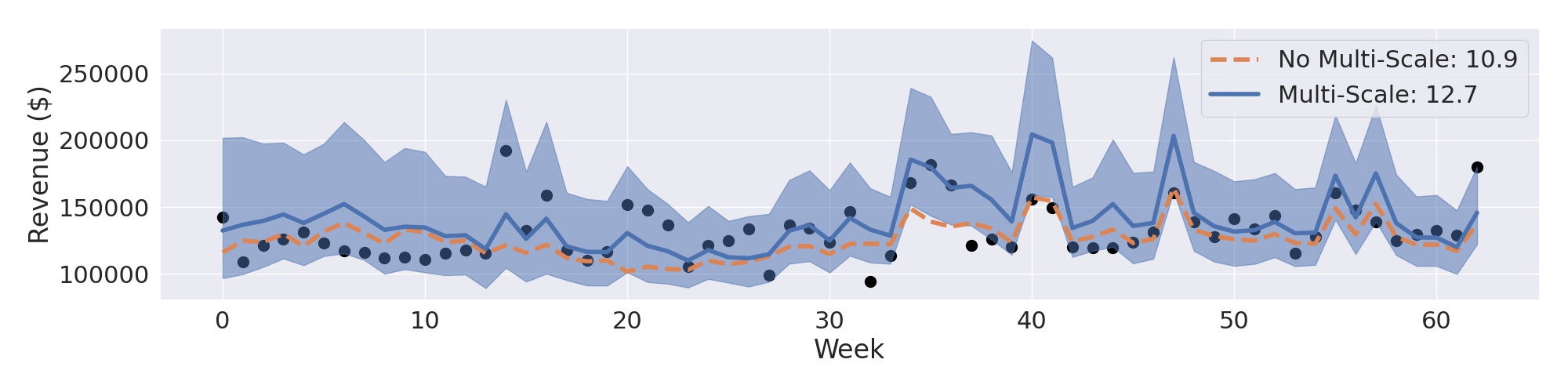}
   \caption{Forecasts - Local Store Group 8.}
\end{subfigure}
\begin{subfigure}[b]{0.95\textwidth}
   \includegraphics[width=1\linewidth]{{"\figdir/revenue/DIV=014-LSG=None-CAT-52-BAKED_SWEET_GOODS-mt-MS-regression-k12"}.png}
   \caption{Regression Effect.}
\end{subfigure}
\caption{12-week ahead forecasts and discount regression effects for the multi-scale and no multi-scale model for the Baked Sweet Goods Category for two LSGs. Format as in \autoref{fig:sugars-main}.}
\label{fig:baked-supp}
\end{figure}

\begin{figure}[htbp!]
\centering
\begin{subfigure}[b]{0.95\textwidth}
   \includegraphics[width=1\linewidth]{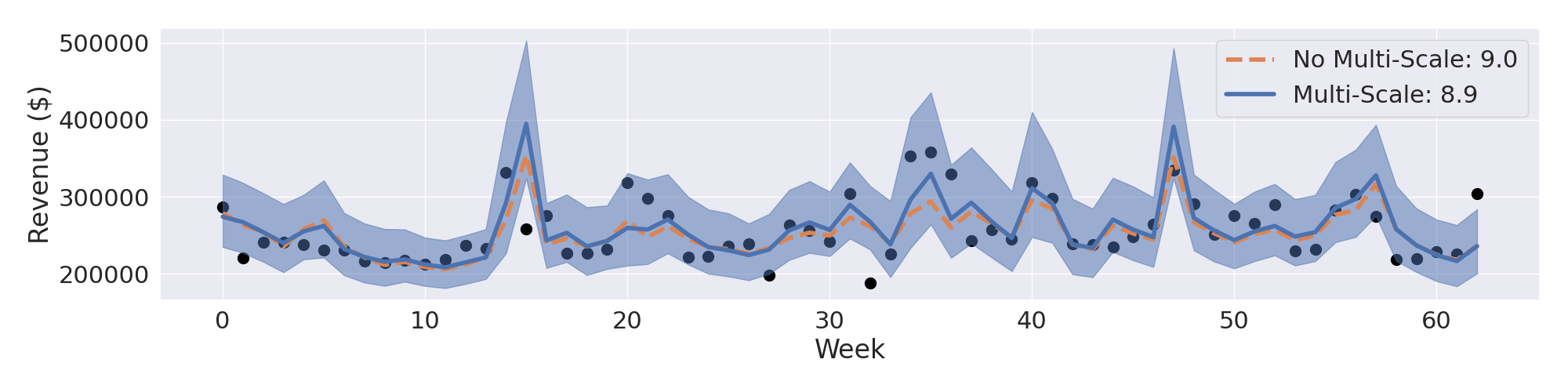}
   \caption{Forecasts - Local Store Group 7.}
\end{subfigure}
\begin{subfigure}[b]{0.95\textwidth}
   \includegraphics[width=1\linewidth]{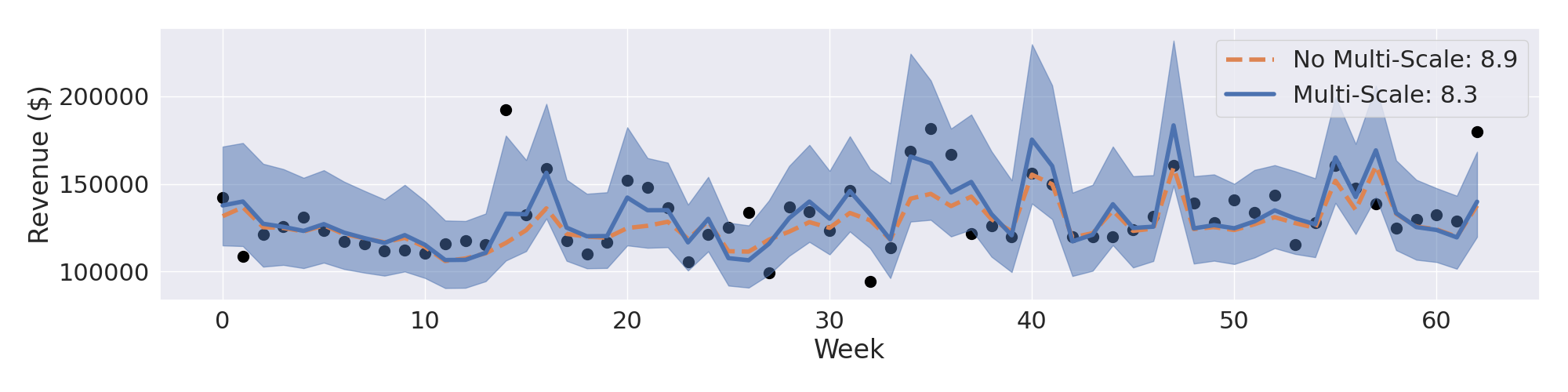}
   \caption{Forecasts - Local Store Group 8.}
\end{subfigure}
\caption{1-week ahead forecasts and discount regression effects for the multi-scale and no multi-scale model for the Baked Sweet Goods Category for two LSGs. Format as in \autoref{fig:sugars-main}.}
\label{fig:baked-supp-k1}
\end{figure}

Forecasts for additional LSGs for each Category discussed in Section~\ref{subsec:results-ms} are given below. The trends discussed in Section~\ref{subsec:results-ms}  hold for Sugars \& Sweeteners (\autoref{fig:sugars-supp}, \autoref{fig:sugars-supp-k1}), Broth/Dry Soup (\autoref{fig:broth-supp}, \autoref{fig:broth-supp-k1}) and Baked Sweet Goods (\autoref{fig:baked-supp}, \autoref{fig:baked-supp-k1}).

\FloatBarrier\newpage

\section*{Additional Joint Pricing and Revenue Forecasting Summaries}\label{app:price}

\begin{figure}[!ht]
\centering
\begin{subfigure}[b]{0.95\textwidth}
   \includegraphics[width=1\linewidth]{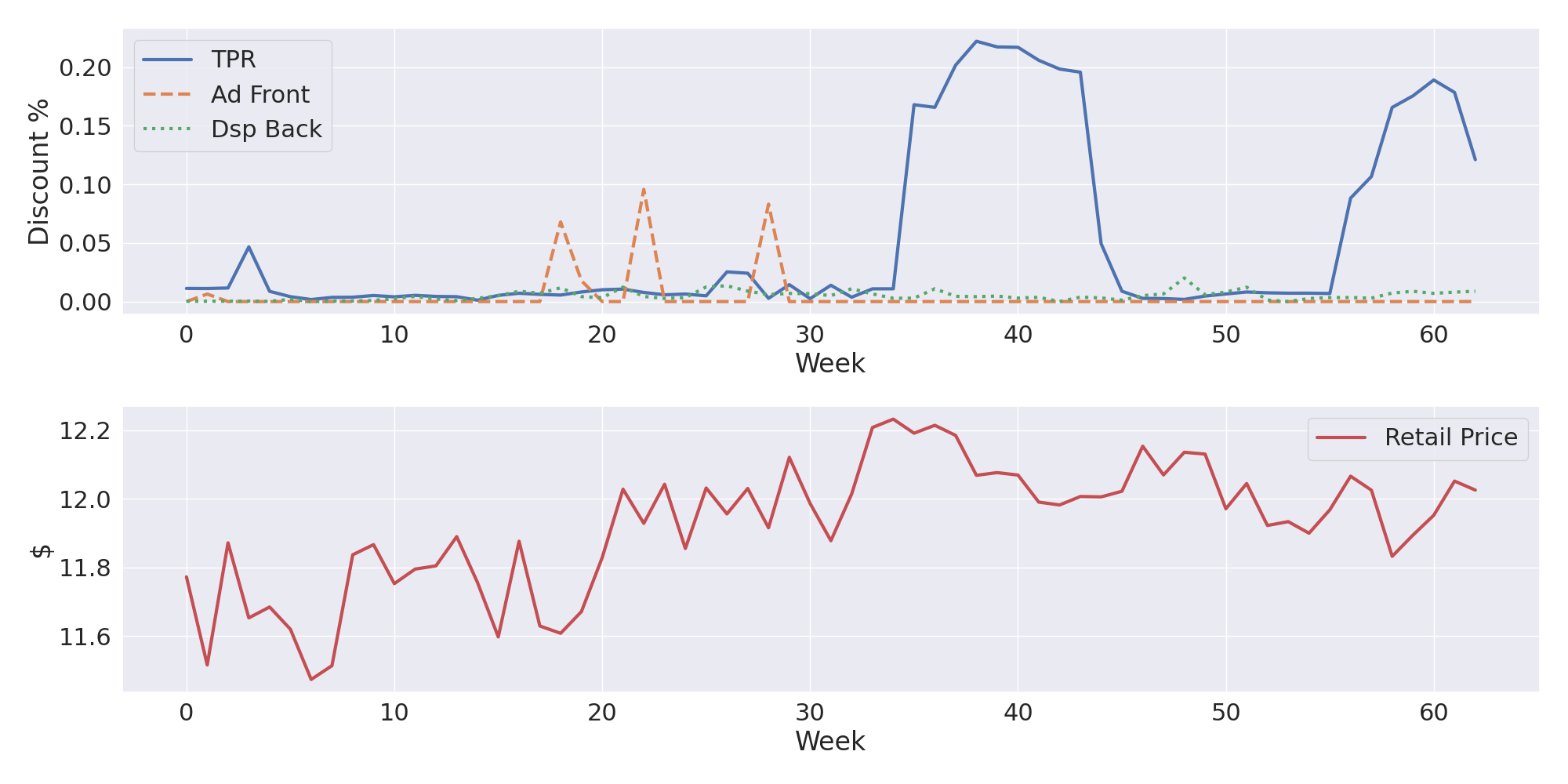}
   \caption{Covariates.}\label{fig:beers-covars}
\end{subfigure}
\begin{subfigure}[b]{0.95\textwidth}
   \includegraphics[width=1\linewidth]{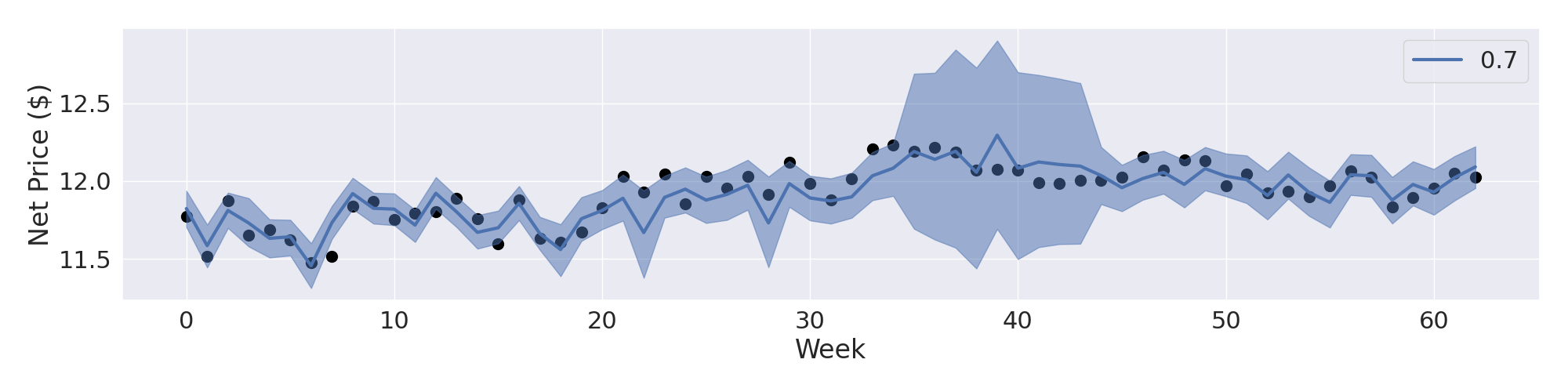}
   \caption{Net Price Forecasts.}\label{fig:beers-net-forecasts}
\end{subfigure}
\caption{(a) Observed discount and pricing covariates for the Net Price model for the Craft/Micro Beers Category in one LSG. (b) Corresponding Net Price forecasts with 90\% credible intervals; the Net Price model includes the covariates given in (a). }
\label{fig:beers-supp}
\end{figure}

Observed covariates and retail price information (\autoref{fig:beers-covars}) for the Craft/Micro Beers example shown in Section~\ref{subsec:improve}.  This Category is rarely discounted and there is some observed price drift over time.  Net Price forecasts for this chosen LSG-Category  pair are given in (\autoref{fig:beers-net-forecasts}).

\FloatBarrier\newpage

\section*{Additional Aspects of Cross-Category Dependence}

\begin{figure}[htbp!]
\centering
\includegraphics[width=0.95\textwidth]{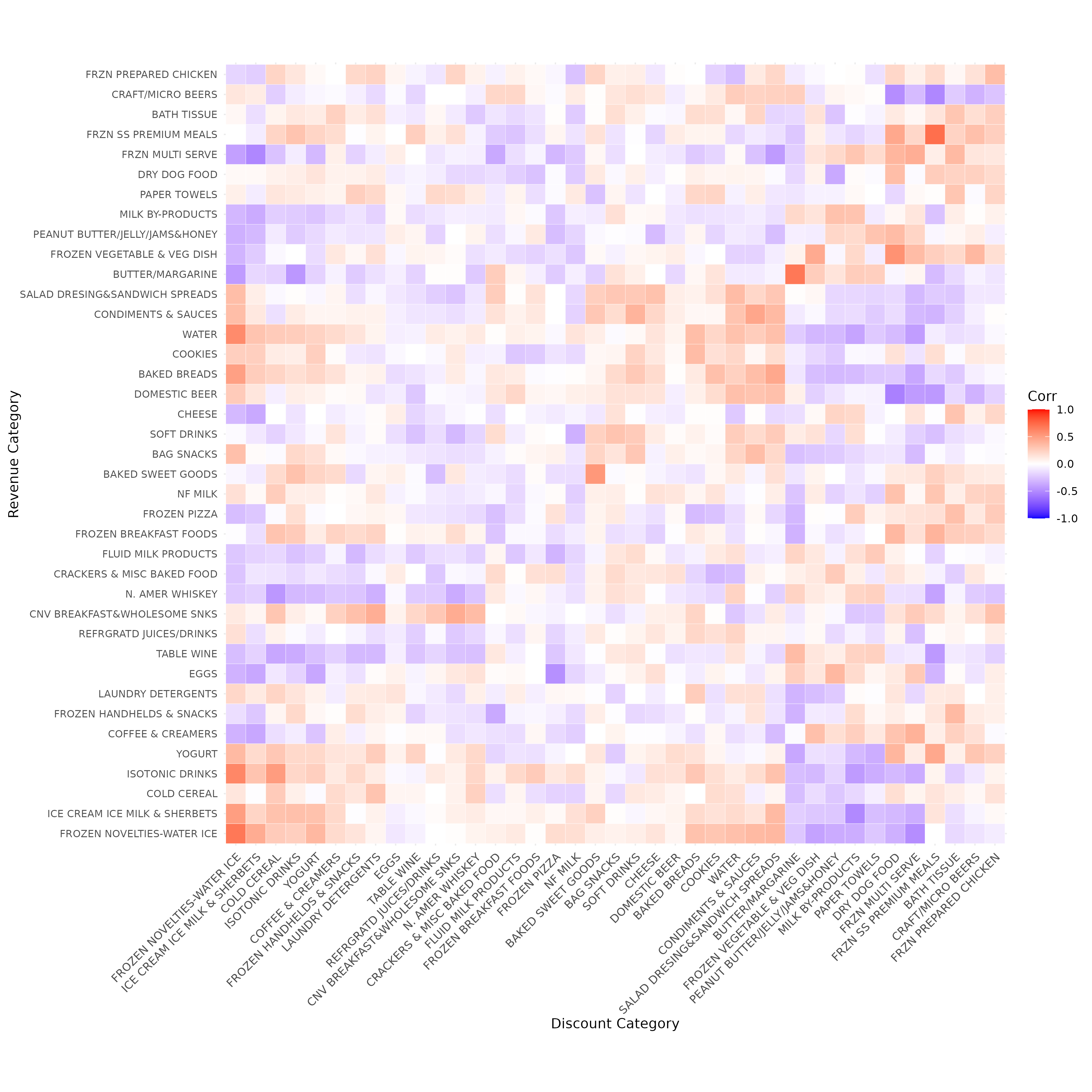}
\caption[Correlations Between Revenue and cross-Category Discounts.]{Heat-map of empirical cross-Category correlations between Revenue and TPR\%.}
\label{fig:correlation_mats_revenue}
\end{figure}

\begin{figure}[htbp!]
\centering
\includegraphics[width=0.95\textwidth]{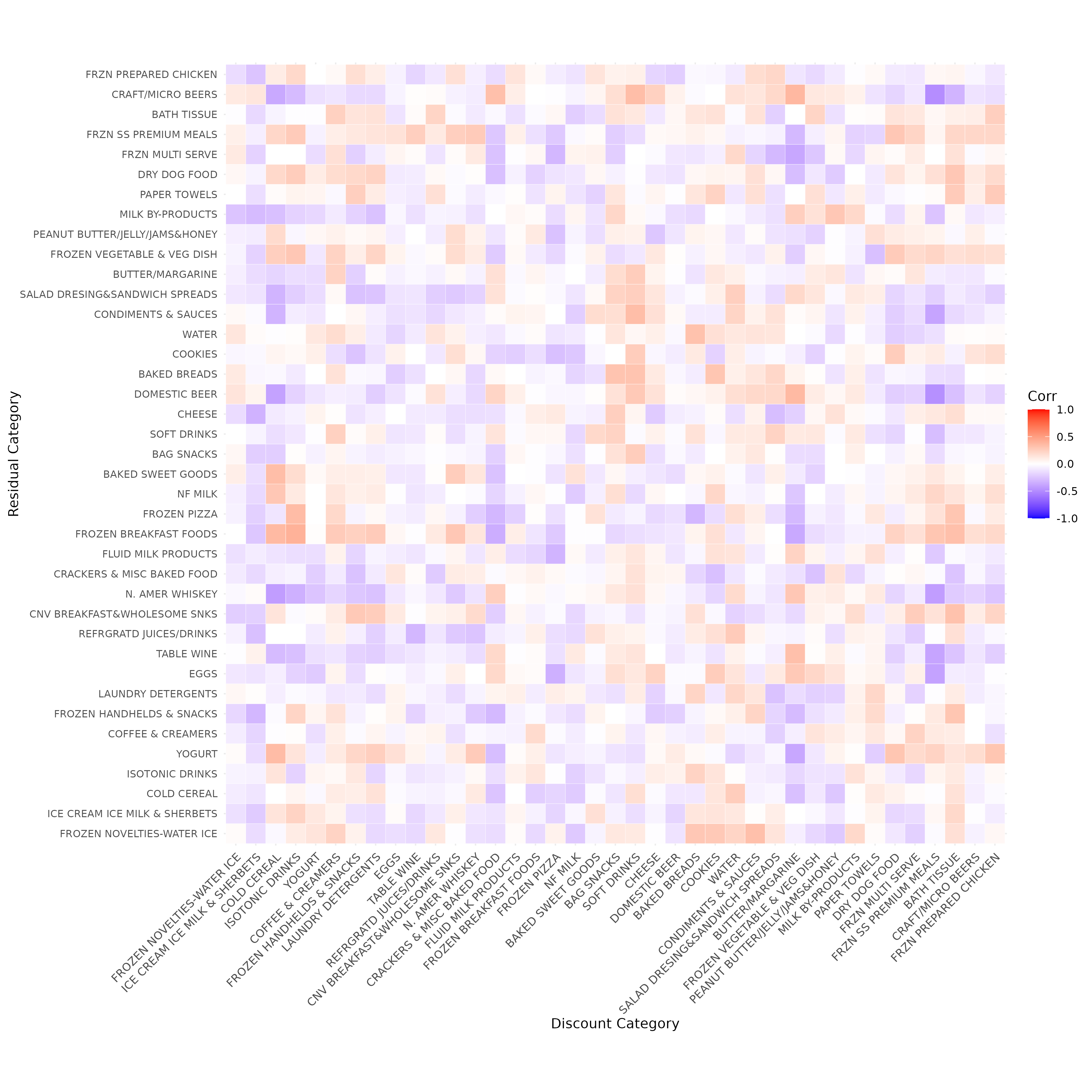}
\caption[Correlations Between 1-week Errors  and cross-Category Discounts.]{Heat-map of empirical cross-Category correlations between realized 1-week ahead Revenue forecast errors and TPR\%. }
\label{fig:correlation_mats_errorsk1}
\end{figure}

Figures~\ref{fig:correlation_mats_revenue} and~\ref{fig:correlation_mats_errorsk1} display 
heat-maps of cross-Category correlations of actual revenue and realized 1-week ahead revenue forecast errors with TPR\% discount. As in the case of 12-week ahead forecast errors (main paper Section~\ref{subsec:cross-cat} and Figure~\ref{fig:correlation_mats_errorsk12}) these are computed over the 52-week forecasting test period and averaged across the 9 LSGs.

\end{document}